\documentclass[prc,twocolumn,showpacs,preprintnumbers,amsmath,amssymb]{revtex4}
\usepackage{rotating}
\usepackage{graphicx}
\usepackage{dcolumn}
\usepackage{amstext}
\usepackage{lscape}
\usepackage{verbatim}
\usepackage[colorlinks,citecolor=blue,bookmarks]{hyperref}
\usepackage{times}
\usepackage{bm}
\usepackage{soul}

\begin{document}

\title{New relativistic effective interaction for finite nuclei, infinite 
nuclear matter and neutron stars}


\author{Bharat Kumar$^{1,2}$}
\author{S. K. Patra$^{1,2}$}
\author{ B. K. Agrawal$^{2,3}$}

\affiliation{\it $^{1}$Institute of Physics, Bhubaneswar-751005, India}
\affiliation{\it $^{2}$Homi Bhabha National Institute, Training School Complex,
Anushakti Nagar, Mumbai 400085, India}
\affiliation{\it $^{3}$Saha Institute of Nuclear Physics, 1/AF,  Bidhannagar, Kolkata - 700064, India.}
\date{\today}

\begin{abstract}
We  carry out the study for finite nuclei, infinite nuclear matter
and neutron star properties with the newly developed relativistic
force named as the Institute Of Physics Bhubaneswar-I(IOPB-I).
Using this force, we calculate the binding energies, charge radii
and neutron-skin thickness for some selected nuclei.  From the ground
state  properties of superheavy nuclei (Z=120), it is noticed  that
considerable shell gaps appear at neutron numbers N=172, 184 and 198,
manifesting  the magicity at these numbers. The low-density behavior of
the equation of state for pure neutron matter is compatible with other
microscopic models. Along with the nuclear symmetry energy, its slope
and curvature parameters at the saturation density are consistent
with those extracted from various experimental data. We calculate
the neutron star properties with the equation of state composed of
nucleons and leptons in $\it beta-equilibrium$ which are in good
agreement with the X-ray observations by Steiner and N\"{a}ttil\"{a}.
Based on the recent observation GW170817 with a quasi-universal relation,
L. Rezzolla {\it et. al.} have set a limit for the maximum mass that can
be supported against gravity by a nonrotating neutron star is in the range
$2.01\pm0.04\lesssim M(M_\odot)$$\lesssim2.16\pm0.03$.  We find that the
maximum mass of the neutron star for the IOPB-I parametrization is 
2.15$M_{\odot}$. The radius and tidal deformability of a {\it canonical}
neutron star mass 1.4$M_\odot$ are 13.2 km and 3.9$\times$10$^{36}$
g cm$^2$ s$^2$ respectively.

\end{abstract}

\maketitle
\section{Introduction}
In the present scenario nuclear physics and nuclear astrophysics are
well described within the self-consistent effective mean field models
\cite{furnstahl97}. These effective theories are not only successful
to describe the properties of finite nuclei  but also explain the
nuclear matter at supra normal densities \cite{aru05}.  Recently,
a large number of nuclear phenomenas have been predicted near the nuclear
drip-lines within  the relativistic and non-relativistic formalisms
\cite{estal01,chab98,sand20}.  Consequently, several experiments are
planed in various laboratories to probe deeper side of the unknown
nuclear territories, {\it i.e.}, the neutron and proton drip-lines.
Among the effective theories, the relativistic mean field (RMF) model is
one of the most successful self-consistent formalism, which has currently
drawn attention for theoretical studies of such systems.

Although, the  construction of the energy density functional for the
RMF model is different than those for the  non-relativistic models,
such as Skyrme \cite{doba84,doba96} and Gogny interactions \cite{dech80},
the obtained results for finite nuclei are in general very close to each
other. The same accuracy in prediction is also valid for the properties
of the neutron stars.  At higher densities, the relativistic effects are
accounted appropriately within the RMF model
\cite{walecka74}. In the RMF model the interactions among nucleons are
described through the exchange of mesons.  These mesons are collectively
taken as effective fields and denoted by classical numbers, which are
the quantum mechanical expectation values. In brief, the RMF formalism
is the relativistic Hartree or Hartree Fock approximations to the one
boson exchange (OBE) theory of nuclear interactions.  In OBE theory, the
nucleons interact with each other by exchange of isovector $\pi-$,
$\rho-$, and $\delta-$mesons and isoscalars like $\eta-$, $\rho-$, and
$\omega-$ mesons. The $\pi-$, and $\eta-$mesons are pseudo-scalar in nature
and do not obey the ground state parity symmetry. In mean-field level,
they do not contribute to the  ground state properties of even nuclei.

The first and simple successful relativistic Lagrangian is formed by
taking only the $\sigma-$, $\omega-$ and $\rho-$ mesons contribution
into account without any non-linear term to the Lagrangian density. This
model predicts an unreasonably large incompressibility $K$ of $\sim 550$
MeV for the infinite nuclear matter at saturation \cite{walecka74}. In order
to lower the value of $K$ to an acceptable range, the self-coupling terms
in sigma meson are included by Boguta and Bodmer \cite{boguta77}. Based
on this Lagrangian density, a large number of parameter sets,
such as NL1 \cite{pg89}, NL2 \cite{pg89}, NL-SH \cite{sharma93},
NL3 \cite{lala97} and NL3$^*$ \cite{nl3s} were  calibrated. The
addition of $\sigma-$meson self-couplings improved the quality of
finite nuclei properties and incompressibility remarkably.  However,
the equation of states at supra-normal densities were quite stiff.
Thus, the addition of vector meson self-coupling is introduced into
the Lagrangian density and different parameter sets are constructed
\cite{bodmer91,gmuca92,sugahara94}. These parameter sets are able to
explain the finite nuclei and nuclear matter properties to a great
extent, but the existence of the Coester-band as well as the 3-body effects
need to be addressed. Subsequently, nuclear physicists also change
their way of thinking and introduced different strategies to improve
the result by designing the density-dependent coupling constants and
effective field theory motivated relativistic mean field (E-RMF) model
\cite{furnstahl97,vretenar00}.

Further, motivated by the effective field theory, Furnstahl {\it
et. al.} \cite{furnstahl97} used all possible couplings up to fourth
order of the expansion, exploiting the naive dimensional analysis (NDA) and
naturalness, obtained  the G1 and G2 parameter sets. In the Lagrangian
density, they considered only the contributions of the isoscalar-isovector
cross-coupling, which has a greater implication on neutron radius and
equation of state (EoS) of asymmetric nuclear matter \cite{pika05}.
Later on it is realized that the contributions of $\delta-$mesons
are also needed to explain certain properties of nuclear phenomena
in extreme conditions \cite{kubis97,G3}. Though the contributions of
the $\delta-$mesons to the  bulk properties are nominal in the normal
nuclear matter, but effects are significant for highly asymmetric
dense nuclear matter.  The $\delta$ meson splits the effective masses
of proton and neutron, which influences the production of $K^{+,-}$ and
$\pi^{+}/\pi^{-}$ in the heavy ion collision (HIC)~\cite{ferini05}. Also,
it increases the proton fraction in the $\beta-$stable matter and modifies
the transport properties of neutron star and heavy ion reactions
\cite{chiu64,bahc65,lati91}.  The source terms for both the $\rho-$
and $\delta-$mesons contain isospin density, but their origins are
different. The $\rho-$meson arises from the asymmetry in the number
density and the evolution of the $\delta-$meson is from the mass asymmetry
of the nucleons.  The inclusion of $\delta-$mesons could influence the
certain physical observables like neutron-skin thickness, isotopic shift,
two neutron separation energy $S_{2n}$, symmetry energy $S(\rho)$,
giant dipole resonance (GDR) and effective mass of the nucleons,
which are correlated with the isovector channel of the interaction.
The density dependence of symmetry energy is  strongly  correlated
with the neutron-skin thickness in heavy nuclei, but till now experiments
have not fixed the accurate value of neutron radius, which is under
consideration for verification in the parity-violating electron-nucleus
scattering experiments~\cite{horowitz01b,vretenar00}.

Recently, the detection of gravitational waves from binary neutron star 
GW170817 is a major breakthrough in astrophysics which is detected for the 
first time by the advanced Laser Interferometer Gravitational- wave 
Observatory (aLIGO) and advanced VIRGO detectors \cite{BNS}. This
detection has certainly posed to be a valuable  guidance to study the
matter under the most extreme conditions.  Inspiraling and coalescing
objects of a binary neutron star result in gravitational waves. Due
to the merger, a compact remnant is remaining whose nature is decided
by two factors {\it i.e} (i) the masses of the inspiraling objects
and (ii) the equation of state of the neutron star matter. For final
state, the formation of either a neutron star or a black hole depends
on the masses and stability of the objects. The chirp mass is measured
very precisely from data analysis  of GW170817 and it is found to be
1.188$^{+0.004}_{-0.002}M_\odot$ for the  90$\%$ credible intervals. It
is suggested that total mass should be 2.74$^{+0.04}_{-0.01}M_\odot$ for
low-spin priors and 2.82$^{+0.47}_{-0.09}M_\odot$ for high-spin priors
\cite{BNS}. Moreover, the maximum mass of nonspinning neutron stars(NSs)
as a function of radius are observed with the highly precise measurements
of $M\approx 2.0M_\odot$. From the observations of gravitational waves,
we can extract information regarding the radii or tidal deformability of
the nonspinning and spinning NSs \cite{flan,tanja,tanja1}. Once we succeed
in getting this information, it is easy to get the neutron star matter
equation of state \cite{latt07,bharat}.

In present paper, we constructed a new parameter set IOPB-I using the
simulated annealing method (SAM) \cite{Agrawal05,Agrawal06,kumar06} and
explore the generic prediction of properties of finite nuclei, nuclear
matter and neutron stars within the E-RMF formalism.  Our new parameter
set yields the considerable shell gap appearing at neutron numbers
N=172, 184 and 198 showing the magicity of these numbers.  The behavior
of the density-dependent symmetry energy of nuclear matter at low and
high densities are examined in detail. The effect of the core EoS on
the mass, radius and tidal deformability of an NS are evaluated using
the static $l=2$ perturbation of a Tolman-Oppenheimer-Volkoff solution.

The paper is organized as follows: In Sec. \ref{theory}, we outline the 
effective field theory motivated relativistic mean field (E-RMF) Lagrangian. 
We outline briefly the equations of motion for finite nuclei and equation of 
states (EoS) for infinite nuclear matter.  In Sec. \ref{parameter}, we 
discuss the strategy of the parameter fitting using the simulated annealing 
method (SAM). After getting the new parameter set IOPB-I, the results on 
binding energy, two neutron separation energy, neutron-skin thickness for 
finite nuclei are discussed in Sec. \ref{result}A. In sub-Sec. \ref{result}B and 
\ref{result}C, the EoS for symmetric and asymmetric matters are presented. 
The mass-radius and tidal deformability of the neutron star obtained by the 
 new parameter set is also discussed in this section. Finally, the summary 
and concluding remarks are given in Sec. \ref{summary}. 

{\it Conventions:} We have taken the value of $G=c=1$ throughout the manuscript.
\section{Formalism}{\label{theory}}

\subsection {Energy density functional and equations of motion}

In this section, we outline briefly the E-RMF Lagrangian \cite{furnstahl97}. 
The beauty of effective Lagrangian is that, one can ignore the basic 
difficulties of the formalism, like renormalization and divergence of the 
system. The model can be used directly by fitting the coupling constants and 
some masses of the mesons.
The  E-RMF Lagrangian has an infinite number of terms with all types of self 
and cross couplings. It is necessary to develop a truncation procedure for
practical use.  
Generally, the meson fields constructed in the Lagrangian are smaller than the 
mass of the nucleon. Their ratio could be used as a truncation scheme as it 
is done in Refs. \cite{furnstahl97,muller96,serot97,estal01} along with the NDA and naturalness properties. The basic nucleon-meson E-RMF Lagrangian 
(with $\delta-$meson, $W\times R$)  up to 4$^{th}$ order with exchange mesons 
like $\sigma-$, $\omega-$, $\rho-$meson and photon $A$ is given as 
\cite{furnstahl97,singh14}: 
\begin{widetext}
\begin{eqnarray}
{\cal E}({r}) & = &  \sum_\alpha \varphi_\alpha^\dagger({r})
\Bigg\{ -i \mbox{\boldmath$\alpha$} \!\cdot\! \mbox{\boldmath$\nabla$}
+ \beta \left[M - \Phi (r) - \tau_3 D(r)\right] + W({r})
+ \frac{1}{2}\tau_3 R({r})
+ \frac{1+\tau_3}{2} A ({r})
\nonumber \\[3mm]
& &
- \frac{i \beta\mbox{\boldmath$\alpha$}}{2M}\!\cdot\!
  \left (f_\omega \mbox{\boldmath$\nabla$} W({r})
  + \frac{1}{2}f_\rho\tau_3 \mbox{\boldmath$\nabla$} R({r}) \right)
  \Bigg\} \varphi_\alpha (r)
  + \left ( \frac{1}{2}
  + \frac{\kappa_3}{3!}\frac{\Phi({r})}{M}
  + \frac{\kappa_4}{4!}\frac{\Phi^2({r})}{M^2}\right )
   \frac{m_s^2}{g_s^2} \Phi^2({r})
\nonumber \\[3mm]
& & \null 
-  \frac{\zeta_0}{4!} \frac{1}{ g_\omega^2 } W^4 ({r})
+ \frac{1}{2g_s^2}\left( 1 +
\alpha_1\frac{\Phi({r})}{M}\right) \left(
\mbox{\boldmath $\nabla$}\Phi({r})\right)^2
 - \frac{1}{2g_\omega^2}\left( 1 +\alpha_2\frac{\Phi({r})}{M}\right)
 \nonumber \\[3mm]
 & &  \null 
 \left( \mbox{\boldmath $\nabla$} W({r})  \right)^2
 - \frac{1}{2}\left(1 + \eta_1 \frac{\Phi({r})}{M} +
 \frac{\eta_2}{2} \frac{\Phi^2 ({r})}{M^2} \right)
  \frac{m_\omega^2}{g_\omega^2} W^2 ({r})
   - \frac{1}{2e^2} \left( \mbox{\boldmath $\nabla$} A({r})\right)^2
   - \frac{1}{2g_\rho^2} \left( \mbox{\boldmath $\nabla$} R({r})\right)^2
    \nonumber \\[3mm]
  & & \null
   - \frac{1}{2} \left( 1 + \eta_\rho \frac{\Phi({r})}{M} \right)
   \frac{m_\rho^2}{g_\rho^2} R^2({r})
   -\Lambda_{\omega}\left(R^{2}(r)\times W^{2}(r)\right)
   +\frac{1}{2 g_{\delta}^{2}}\left( \mbox{\boldmath $\nabla$} D({r})\right)^2
   +\frac{1}{2}\frac{ {  m_{\delta}}^2}{g_{\delta}^{2}}\left(D^{2}(r)\right)\;,
\label{eq1}
\end{eqnarray}
\end{widetext}
where $\Phi$, $W$, $R$, $D$ and $A$ are the fields, $g_\sigma$, $g_\omega$, $g_\rho$, $g_\delta$, 
$\frac{e^2}{4\pi}$  are the coupling constants and $m_\sigma$, $m_\omega$, $m_\rho$ and $m_\delta$
are the masses for $\sigma$, $\omega$, $\rho$, $\delta$ mesons and photon, respectively.

Now, our aim is to solve the field equations for the baryons and mesons 
(nucleon, $\sigma$, $\omega$, $\rho$, $\delta$) using the variational principle. We obtained the mesons equation of motion using the equation 
	$\left(\frac{\partial\mathcal{E}}{\partial\phi_i}\right)_{\rho=constant}=0.$
%
The single particle energy for the nucleons is obtained by using the Lagrange
multiplier $\varepsilon_\alpha$, which is the energy eigenvalue of the
Dirac equation constraining the normalization condition 
$\sum_\alpha\varphi_\alpha^\dagger({r}) \varphi_\alpha({r})=1 $ \cite{furn87}.  
The Dirac equation for the wave function $\varphi_\alpha({r})$ becomes
\begin{eqnarray}
\frac{\partial}{\partial\varphi_\alpha^\dagger({r})}\Bigg[{\cal E}{({r})}-
\sum_\alpha\varphi_\alpha^\dagger({r}) \varphi_\alpha({r})\Bigg]= 0,
\end{eqnarray}
{\it i.e.}
\begin{widetext}
\begin{eqnarray}
\Bigg\{-i \mbox{\boldmath$\alpha$} \!\cdot\! \mbox{\boldmath$\nabla$}
& + & \beta [M - \Phi(r) - \tau_3 D(r)] + W(r)
  +\frac{1}{2} \tau_3 R(r) + \frac{1 +\tau_3}{2}A(r)
\nonumber \\[3mm]
& & \null  
\left.
   -  \frac{i\beta \mbox{\boldmath$\alpha$}}{2M}
   \!\cdot\! \left [ f_\omega \mbox{\boldmath$\nabla$} W(r)
    + \frac{1}{2}f_{\rho} \tau_3 \mbox{\boldmath$\nabla$} R(r) \right]
     \right\} \varphi_\alpha (r) =
     \varepsilon_\alpha \, \varphi_\alpha (r) \,.
     \label{eq2}
     \end{eqnarray}
\end{widetext}
The mean-field equations for $\Phi$, $W$, $R$, $D$ and $A$ are given by
\begin{widetext}
\begin{eqnarray}
-\Delta \Phi(r) + m_s^2 \Phi(r)  & = &
g_s^2 \rho_s(r)
-{m_s^2\over M}\Phi^2 (r)
\left({\kappa_3\over 2}+{\kappa_4\over 3!}{\Phi(r)\over M}
\right )
\nonumber  \\[3mm]
& & \null
+{g_s^2 \over 2M}
\left(\eta_1+\eta_2{\Phi(r)\over M}\right)
{ m_\omega^2\over  g_\omega^2} W^2 (r)
+{\eta_{\rho} \over 2M}{g_s^2 \over g{_\rho}^2}
{ m_\rho^2 } R^2 (r)
\nonumber  \\[3mm]
& & \null
+{\alpha_1 \over 2M}[
(\mbox{\boldmath $\nabla$}\Phi(r))^2
+2\Phi(r)\Delta \Phi(r) ]
+ {\alpha_2 \over 2M} {g_s^2\over g_\omega^2}
(\mbox{\boldmath $\nabla$}W(r))^2 \,,
 \label{eq3}  \\[3mm]
-\Delta W(r) +  m_\omega^2 W(r)  & = &
g_\omega^2 \left( \rho(r) + \frac{f_\omega}{2} \rho_{\rm T}(r) \right)
-\left( \eta_1+{\eta_2\over 2}{\Phi(r)\over M} \right ){\Phi(r)
\over M} m_\omega^2 W(r)
\nonumber  \\[3mm]
& & \null
-{1\over 3!}\zeta_0 W^3(r)
+{\alpha_2 \over M} [\mbox{\boldmath $\nabla$}\Phi(r)
\cdot\mbox{\boldmath $\nabla$}W(r)
+\Phi(r)\Delta W(r)]
\nonumber\\
& &
-2\;\Lambda_{\omega}{g_{\omega}}^2 {R^{2}(r)} W(r) \,,
\label{eq4}  \\[3mm]
-\Delta R(r) +  m_{\rho}^2 R(r)  & = &
{1 \over 2 }g_{\rho}^2 \left (\rho_{3}(r) +
{1 \over 2 }f_{\rho}\rho_{\rm T,3}(r)
\right ) -  \eta_\rho {\Phi (r) \over M }m_{\rho}^2 R(r)
-2\;\Lambda_{\omega}{g_{\rho}}^{2} R(r) {W^{2}(r)} \,,
\label{eq5}  \\[3mm]
-\Delta A(r)   & = &
e^2 \rho_{\rm p}(r)    \,, 
\label{eq6} \\[3mm]
-\Delta D(r)+{m_{\delta}}^{2} D(r) & = &
g_{\delta}^{2}\rho_{s3} \;,
\label{eq7}
\end{eqnarray}
\end{widetext}
where the baryon, scalar, isovector, proton and tensor densities are
\begin{eqnarray}
\rho(r) & = &
\sum_\alpha \varphi_\alpha^\dagger(r) \varphi_\alpha(r) \,
\nonumber\\
	&=&\rho_{p}(r)+\rho_{n}(r)
\nonumber\\
&= &
\frac{2}{(2\pi)^{3}}\int_{0}^{k_{p}}d^{3}k
+\frac{2}{(2\pi)^{3}}\int_{0}^{k_{n}}d^{3}k,
\label{den_nuc} 
\end{eqnarray}

\begin{eqnarray}
\rho_s(r) & = &
\sum_\alpha \varphi_\alpha^\dagger(r) \beta \varphi_\alpha(r) 
\nonumber \\ 
	&=&\rho_{s p}(r) +  \rho_{s n}(r)
\nonumber \\
&=&\sum_{\alpha} \frac{2}{(2\pi)^3}\int_{0}^{k_{\alpha}} d^{3}k 
\frac{M_{\alpha}^{\ast}} {(k^{2}_\alpha+M_{\alpha}^{\ast 2})^{\frac{1}{2}}},
\label{scalar_density} 
\end{eqnarray}

\begin{eqnarray}
\rho_3 (r) & = &
\sum_\alpha \varphi_\alpha^\dagger(r) \tau_3 \varphi_\alpha(r) \nonumber \\
&=& \rho_{p} (r) -  \rho_{n} (r),
\label{isovect_density}
\end{eqnarray}
\begin{eqnarray}
\rho_{s3} (r) & = &
\sum_\alpha \varphi_\alpha^\dagger(r) \tau_3 \beta \varphi_\alpha(r) 
\nonumber \\
&=& \rho_{ps} (r) -  \rho_{ns} (r)
\label{isoscal_density}
\end{eqnarray}
\begin{eqnarray}
\rho_{\rm p}(r) & = &
\sum_\alpha \varphi_\alpha^\dagger(r) \left (\frac{1 +\tau_3}{2}
\right)  \varphi_\alpha(r),
\label{charge_density} 
\end{eqnarray}
\begin{eqnarray}
\rho_{\rm T}(r) & = &
\sum_\alpha \frac{i}{M} \mbox{\boldmath$\nabla$} \!\cdot\!
\left[ \varphi_\alpha^\dagger(r) \beta \mbox{\boldmath$\alpha$}
\varphi_\alpha(r) \right] \,,
\label{tensor_dens} 
\end{eqnarray}
and
\begin{eqnarray}
\rho_{\rm T,3}(r) & = &
\sum_\alpha \frac{i}{M} \mbox{\boldmath$\nabla$} \!\cdot\!
\left[ \varphi_\alpha^\dagger(r) \beta \mbox{\boldmath$\alpha$}
\tau_3 \varphi_\alpha(r) \right].  
\label{isotensor_dens} 
\end{eqnarray}
Here $k_{\alpha}$ is the nucleon's  Fermi momentum and the summation is 
over all the occupied states. The nucleons and mesons are composite particles 
and their vacuum polarization effects have been neglected. Hence, the 
negative-energy states do not contribute to the densities and current 
\cite{pg89}. In the 
fitting process, the coupling constants of the effective Lagrangian are
 determined from a set of experimental data which takes into account the large 
part of the vacuum polarization effects in the {\it no-sea approximation}. 
It is clear that the {\it no-sea approximation} is essential to determine 
the stationary solutions of the relativistic mean-field equations which 
describe the ground-state properties of the nucleus. The Dirac sea holds the 
negative energy eigenvectors of the Dirac Hamiltonian which is different 
for different nuclei. Thus, it depends on the specific solution of the set of 
nonlinear RMF equations. The Dirac spinors can be expanded in terms of vacuum 
solutions which form a complete set of plane wave functions in spinor space. 
This set will be complete when the states with negative energies are the part 
of the positive energy states and create the Dirac sea of the vacuum.

The effective mass of proton $M_{p}^{\ast}$ and neutron $M_{n}^{\ast}$ are written as 
\begin{equation}
M_{p}^{\ast}=M- \Phi (r) - D({r}),
\end{equation}
\begin{equation} 
M_{n}^{\ast}=M- \Phi (r) + D({r}).
\end{equation}
The vector potential is
\begin{equation}
V(r)=g_{\omega}V_{0}(r)+\frac{1}{2} g_{\rho}\tau_{3}b_{0}(r)
+e\frac{(1-\tau_3)}{2}A_0(r). 
\end{equation}
The set of coupled differential equations are solved self-consistently to describe the
ground state properties of finite nuclei. In the fitting procedure, we used the experimental
data of binding energy (BE) and charge radius $r_{ch}$ for a set of spherical nuclei
($^{16}$O, $^{40}$Ca, $^{48}$Ca, $^{68}$Ni, $^{90}$Zr, $^{100,132}$Sn and  $^{208}$Pb). 
The total binding energy is obtained by 
\begin{eqnarray}
E_{total} &=& E_{part}+E_{\sigma}+E_{\omega}+E_{\rho}
\nonumber\\   
&&
	+E_{\delta}+E_{\omega\rho}+E_{c}+E_{pair}+E_{c.m.},
\end{eqnarray}
where $E_{part}$ is the sum of the single particle energies of the nucleons and 
$E_{\sigma}$, $E_{\omega}$, $E_{\rho}$, $E_{\delta}$, $E_{c}$ are the 
contributions of the respective mesons and Coulomb fields. The pairing 
$E_{pair}$ and the center of mass motion $E_{cm}=\frac{3}{4}\times41A^{-1/3}$ MeV 
energies are also taken into account \cite{elliott55,negele70,estal01}.

The pairing correlation plays a distinct role in open-shell 
nuclei \cite{greiner72,ring80}. The effect of pairing correlation is markedly 
seen with the increase in mass number A. Moreover, it helps in understanding 
the deformation of medium and heavy nuclei. It has a lean 
effect on both bulk and single particle properties of lighter mass nuclei
because of the availability of limited pairs near the Fermi surface. We take
the case of T=1 channel of pairing correlation i.e, pairing between proton-
proton and neutron-neutron. The pairs of nucleons are invariant under time 
reversal symmetry when the pairing interaction $v_{pair}$ has non-zero matrix 
elements:
\begin{eqnarray}
\langle \alpha_2 \overline{\alpha_2} \vert v_{pair}\vert\alpha_1 \overline{\alpha_1}\rangle = -G ,
\end{eqnarray}
where $\alpha=\vert nljm\rangle$ and $\overline{\alpha}=\vert nlj-m\rangle$ 
 (with $G>0$ and $m>0$) are the quantum states.

A nucleon of quantum states $\vert nljm\rangle$ pairs with another nucleon 
having same $I_z$ value with quantum states $\vert nlj-m\rangle$,
 since it is the time reversal partner of the other. In both nuclear and atomic
domains, the ideology of BCS pairing is the same. The even-odd mass staggering 
of isotopes was the first evidence of its kind for the pairing energy. 
Considering the mean-field formalism the violation of the particle number is
seen only due to the pairing correlation. We find terms such as
 $\varphi^{\dagger}\varphi$ (density) in the RMF Lagrangian density but we put an embargo on terms
of the form $\varphi^{\dagger}\varphi^{\dagger}$ or $\varphi\varphi$ since it 
violates the particle number conservation. Thus, we affirm that BCS 
calculations have been carried out by constant gap or constant force approach 
externally in the RMF model \cite{gambhir90, reinhard86,toki94}. In our work, 
we  consider seniority type interaction as a tool by taking a constant value of 
G for pairs of the active pair shell.

The above approach does not go well for nuclei away from the stability
line  because in the present case, with the increase in the number of 
neutrons or protons the corresponding Fermi level goes to zero and 
the number of available levels above it minimizes. To complement this
situation we see that the particle-hole and pair excitations reach the 
continuum. In Ref. \cite{doba84} we notice that if we make the BCS calculation 
using the quasiparticle state as in HFB calculation, then the BCS binding 
energies are coming out to be very close to the HFB, but rms radii (i.e the single-particle wave functions) greatly depend on the size of the box where the 
calculation is done. This is because of the unphysical neutron (proton) gas
in the continuum where wavefunctions are not confined in a region.
The above shortcomings of the BCS approach can be improved by 
means of the so-called quasibound states, {\it i.e}, states bound because of 
their own centrifugal barrier (centrifugal-plus-Coulomb barrier for protons)
\cite{estal01,meyer98,liotta20}. Our calculations are done by confining the 
available space to one harmonic oscillator shell each above and below the Fermi 
level to exclude the unrealistic pairing of highly excited states in the continuum \cite{estal01}.


\subsection{Nuclear Matter Properties}{\label{nuclear-matter}}

\begin{center}
\item{ \bf 1. Energy and pressure density}
\end{center}
In a static, infinite, uniform and isotropic nuclear matter, all
the gradients of the fields in Eqs.(\ref{eq3})-(\ref{eq7}) vanish. By the 
definition of infinite nuclear matter, the electromagnetic interaction is also 
neglected.
The expressions for energy density and pressure for such a system is obtained
from the energy-momentum tensor \cite{walecka86}:
\begin{eqnarray}
T_{\mu \nu}=\sum_{i}{\partial_{\nu} \phi_{i}} 
\frac{ {\partial \cal L}} {\partial \left({\partial^{\mu} \phi_{i}}\right)} 
-g_{\mu \nu}{\cal L}.
\end{eqnarray}
The zeroth component of the energy-momentum tensor $<T_{00}>$ gives the 
energy density and the third component $<T_{ii}>$ compute the pressure of 
the system \cite{singh14}:
\begin{widetext}
\begin{eqnarray}\label{eqn:eos1}
{\cal{E}} & = &  \frac{2}{(2\pi)^{3}}\int d^{3}k E_{i}^\ast (k)+\rho  W+
\frac{ m_{s}^2\Phi^{2}}{g_{s}^2}\Bigg(\frac{1}{2}+\frac{\kappa_{3}}{3!}
\frac{\Phi }{M} + \frac{\kappa_4}{4!}\frac{\Phi^2}{M^2}\Bigg)
\nonumber\\
&&
 -\frac{1}{2}m_{\omega}^2\frac{W^{2}}{g_{\omega}^2}\Bigg(1+\eta_{1}\frac{\Phi}{M}+\frac{\eta_{2}}{2}\frac{\Phi ^2}{M^2}\Bigg)-\frac{1}{4!}\frac{\zeta_{0}W^{4}}
	{g_{\omega}^2}+\frac{1}{2}\rho_{3} R
 \nonumber\\
 &&
-\frac{1}{2}\Bigg(1+\frac{\eta_{\rho}\Phi}{M}\Bigg)\frac{m_{\rho}^2}{g_{\rho}^2}R^{2}-\Lambda_{\omega}  (R^{2}\times W^{2})
+\frac{1}{2}\frac{m_{\delta}^2}{g_{\delta}^{2}}\left(D^{2} \right),
	\label{eq20}
\end{eqnarray}
\end{widetext}


\begin{widetext}
\begin{eqnarray}\label{eqn:eos2}
P & = &  \frac{2}{3 (2\pi)^{3}}\int d^{3}k \frac{k^2}{E_{i}^\ast (k)}-
\frac{ m_{s}^2\Phi^{2}}{g_{s}^2}\Bigg(\frac{1}{2}+\frac{\kappa_{3}}{3!}
\frac{\Phi }{M}+ \frac{\kappa_4}{4!}\frac{\Phi^2}{M^2}  \Bigg)
\nonumber\\
& &
 +\frac{1}{2}m_{\omega}^2\frac{W^{2}}{g_{\omega}^2}\Bigg(1+\eta_{1}\frac{\Phi}{M}+\frac{\eta_{2}}{2}\frac{\Phi ^2}{M^2}\Bigg)+\frac{1}{4!}\frac{\zeta_{0}W^{4}}{g_{\omega}^2}
  \nonumber\\
& &
+\frac{1}{2}\Bigg(1+\frac{\eta_{\rho}\Phi}{M}\Bigg)\frac{m_{\rho}^2}{g_{\rho}^2}R^{2}+\Lambda_{\omega} (R^{2}\times W^{2})
-\frac{1}{2}\frac{m_{\delta}^2}{g_{\delta}^{2}}\left(D^{2}\right),
	\label{eq21}
\end{eqnarray}
\end{widetext}
where $E_{i}^\ast(k)$=$\sqrt {k^2+{M_{i}^\ast}^2} \qquad  (i= p,n)$
is the energy and $k$ is the momentum of the nucleon. In the context of 
density functional theory, it is possible to parametrize the exchange and correlation effects 
through local potentials (Kohn--Sham potentials), as long as those 
contributions are small enough \cite{Ko65}. The Hartree values control the 
dynamics in the relativistic Dirac-Br\"uckner-Hartree-Fock (DBHF) calculations. Therefore, the local meson fields in the RMF formalism can be interpreted as 
Kohn-Sham potentials and in this sense equations (\ref{eq2}-\ref{eq7}) include 
effects beyond the Hartree approach through the non-linear couplings \cite{furnstahl97}.

\begin{center}
 {\bf 2. Symmetry Energy}
\end{center}

The binding energy per nucleon ${\cal E}/A$=$e(\rho, \alpha)$ can be written 
in the parabolic form of asymmetry parameter $\alpha\left(=\frac{\rho_n-\rho_p}{\rho_n+\rho_p}\right)$:
\begin{eqnarray}
e(\rho, \alpha)=\frac{{\cal E}}{\rho_{B}} - M =
{e}(\rho) + S(\rho) \alpha^2 + 
{\cal O}(\alpha^4) ,
\end{eqnarray}
where ${e}(\rho)$ is energy density of the symmetric nuclear matter (SNM) ($\alpha$ = 0) 
and $S(\rho)$ is defined as the symmetry energy of the system:
\begin{eqnarray}
S(\rho)=\frac{1}{2}\left[\frac{\partial^2 {e}(\rho, \alpha)}
{\partial \alpha^2}\right]_{\alpha=0}.
\end{eqnarray}
The isospin asymmetry arises due to the difference in densities and masses 
of the neutron and proton. The density type isospin asymmetry 
is taken care by $\rho-$meson (isovector-vector meson) and mass asymmetry 
by $\delta-$meson (isovector - scalar meson). The general expression for 
symmetry energy $S(\rho)$ is a combined expression of $\rho-$ and $\delta-$mesons, which is defined as~\cite{matsui81,kubis97,estal01,roca11}:
\begin{eqnarray}
S(\rho)=S^{kin}(\rho) + S^{\rho}({\rho})+S^{\delta}
(\rho),
\end{eqnarray}
with
\begin{eqnarray}
S^{kin}(\rho)=\frac{k_F^2}{6E_F^*},\;
S^{\rho}({\rho})=\frac{g_{\rho}^2\rho}{8m_{\rho}^{*2}}
\end{eqnarray}
and
\begin{eqnarray}
S^{\delta}(\rho)=-\frac{1}{2}\rho \frac{g_\delta^2}{m_\delta^2}\left(
\frac{M^*}{E_F}\right)^2 u_\delta \left(\rho,M^*\right).
\end{eqnarray}
The last function $u_\delta$ is from the discreteness of the Fermi momentum. 
This momentum is quite large in nuclear matter and can be treated as a 
continuum and continuous system.
The function $u_\delta$ is defined as:
\begin{eqnarray} 
u_\delta \left(\rho,M^*\right)=\frac{1}{ 1+ 3 \frac{g_\delta^2}{m_\delta^2}
\left(\frac{\rho^s}{M^*}-\frac{\rho}{E_F}\right)}.
\end{eqnarray}
In the limit of continuum, the function $u_\delta \approx 1$. 
The whole symmetry energy ($S^{kin}+S^{pot}$) arises from 
$\rho-$ and $\delta-$mesons is given as:
\begin{eqnarray}{\label{eqn:sym}}
S(\rho)=\frac{k_F^2}{6E_F^*}+\frac{g_{\rho}^2\rho}{8{m_{\rho}^{*2}}}  
-\frac{1}{2}\rho \frac{g_\delta^2}{m_\delta^2}\left(\frac{M^*}{E_F}\right)^2,
\end{eqnarray}
where $E_F^*$ is the Fermi energy and $k_F$ is the
Fermi momentum. 
The mass of the $\rho$-meson modified, because of the cross-coupling of 
$\rho-\omega$ fields and is given by
\begin{eqnarray}
m_{\rho}^{*2}=\left(1+\eta_{\rho}\frac{\Phi}{M}\right)m_{\rho}^2
+2g_{\rho}^2(\Lambda_\omega W^2).
\end{eqnarray}
The cross-coupling of isoscalar-isovector mesons ($\Lambda_\omega$)  
modifies the density dependence of $S(\rho)$ without affecting the 
saturation properties of the symmetric nuclear matter 
(SNM)~\cite{singh13,horowitz01b,horowitz01a}.
In the numerical calculation, the coefficient of symmetry energy $S(\rho)$ 
is obtained by the energy difference of symmetric and pure neutron matter 
at saturation. In our calculation, we have taken the isovector channel into 
account to make the new parameters, which incorporate the currently existing 
experimental observations and predictions are done keeping in mind some future 
aspects of the model. The symmetry energy can be expanded as a Taylor series 
around the saturation density $\rho_0$ as:
\begin{eqnarray}
S(\rho)=J + L{\cal Y} 
+ \frac{1}{2}K_{sym}{\cal Y}^2 +\frac{1}{6}Q_{sym}{\cal Y}^3 + {\cal O}[{\cal Y}^4],
	\label{eq30}
\end{eqnarray} 
where $J=S(\rho_0)$ is the symmetry energy at saturation and 
${\cal Y} = \frac{\rho-\rho_0}{3\rho_0}$. The coefficients $L(\rho_0)$, 
$K_{sym}(\rho_0)$, and $Q_{sym}$ are defined as:
\begin{eqnarray}
	L=3\rho\frac{\partial S(\rho)}{\partial {\rho}}\bigg{|}_{ \rho=\rho_0},\;
\end{eqnarray}
\begin{eqnarray}
	K_{sym}=9\rho^2\frac{\partial^2 S(\rho)}{\partial {\rho}^2}\bigg{|}_{ {\rho=\rho_0}=0},\;
\end{eqnarray}
\begin{eqnarray}
        Q_{sym}=27\rho^3\frac{\partial^3 S(\rho)}{\partial {\rho}^3}\bigg{|}_{ {\rho=\rho_0}=0}.\;
\end{eqnarray}
Similarly, we obtain the asymmetric nuclear matter (ANM)  
incompressiblity as $K(\alpha)=K+K_\tau \alpha^2+{\cal O}(\alpha^4)$ and
$K_\tau$ is given by \cite{chen09}
\begin{eqnarray}
	K_{\tau}=K_{sym}-6L-\frac{Q_{0}L}{K},\;
	\label{eq34}
\end{eqnarray}
where $Q_{0}=27\rho^3\frac{\partial^3 (\cal{E}/\rho)}{\partial {\rho}^3}$ in SNM.

Here, $L$ is the slope and $K_{sym}$ represents the curvature of $S(\rho)$ at 
saturation density. A large number of investigations have been made to fix the 
values of $J$, $L$ and $K_{sym}$ \cite{singh13,tsang12,dutra12,xu10,newton11,steiner12,fattoyev12}. The density dependence of symmetry energy is a key quantity 
to control the properties of both finite nuclei and infinite nuclear matter 
\cite{roca09}. Currently, the available information on symmetry energy 
$J=31.6\pm2.66$ MeV and its slope $L=58.9\pm16$ MeV at saturation density are 
obtained by various astrophysical observations \cite{li13}.  
Till date, the precise values of $J,L$ and the neutron radii for finite
nuclei are not known experimentally, it is essential to discuss the behavior 
of the symmetry energy as a function of density in our new parameter set. 

\section{PARAMETER FITTING}{\label{parameter}}

The simulated annealing method (SAM) is used to determine the parameters used 
in the Lagrangian density \cite{kirkpatrick83,press92}.  The SAM is useful in 
the global minimization technique, {\it i.e.},  it gives accurate results when 
there exists a global minimum within several local minima. Usually, this 
procedure is used in a system when the number of parameters are more than the 
observables \cite{kirkpatrick84,ingber89,cohen94}.
In this simulation method, the system stabilizes, when the temperature 
$T$ (a variable which controls the energy of the system) goes down \cite{Agrawal05,Agrawal06,kumar06}. Initially, 
the nuclear system is put at a high temperature (highly unstable) and then 
allowed to cool down slowly so that it is stabilized in a very smooth way and 
finally reaches the frozen temperature (stable or systematic system). The 
variation of $T$ should be very small near the stable state. The 
$\chi^2 =\chi^2(p_{1},\dots\dots\dots,p_{N}) $ 
values of the considered systems are minimized (least-square fit), which is 
governed by the model parameters $p_i$. The general expression of the $\chi^2$ can be 
given as:
 \begin{eqnarray}
\chi^{2}= \frac{1}{N_{d}-N_{p}} \sum_{i=1}^{N_{d}}
\left(\frac{M_{i}^{exp}-M_{i}^{th}}{\sigma_{i}}\right)^2.
\label{eq:chi2}\end{eqnarray}  
Here, $N_{d}$ and $N_{p}$ are the number of experimental data points and
the fitting parameters, respectively. The experimental and theoretical
values of the observables are denoted by $M_{i}^{exp}$ and $M_{i}^{th}$,
respectively.  The $\sigma_{i}$'s are the adopted errors \cite{dob14}.
The adopted errors are composed of three components, namely, the
experimental, numerical and theoretical errors \cite{dob14}. As the name
suggests, the experimental errors are associated with the measurements,
numerical and the theoretical errors are associated with the  numerics
and the shortcomings of the nuclear model employed, respectively. In
principle, there exists some arbitrariness in choosing the values of
$\sigma_i$  which is partially responsible for the proliferation of
the mean-field models.  Only guidance  available from the statistical
analysis is that the chi-square per degree of freedom (Eq. \ref{eq:chi2})
should be close to the unity.
In the present calculation, we have  used some selected fit data for
binding energy and root mean square radius of charge distribution for
some selected nuclei and the associated adopted errors on them
\cite{klup09}.



In our calculations, we have built a new parameter set IOPB-I and
analyzed their effects for finite and infinite nuclear systems. Thus,
we performed an overall fit with 8 parameters, where the nucleons as well as 
the masses of the two vector mesons in free space are fixed at their 
experimental values {\it i.e.}, $M=939$ MeV, $m_{\omega}=782.5$ MeV 
and $m_{\rho}=763.0$ MeV.  The effective nucleon mass can be used as 
a nuclear matter
constraint at the saturation density $\rho_0$ along with other empirical
values like incompressibility, binding energy per nucleon and asymmetric
parameter $J$.  While fitting the parameter, the value of effective
nucleon mass $M^{*}/M$, nuclear matter incompressibility $K$ and
symmetry energy coefficient $J$ are constrained within 0.50-0.90, 220-260
MeV and 28-36 MeV, respectively.  The minimum $\chi^{2}$ is obtained by
simulated annealing method \cite{ Agrawal05,Agrawal06,kumar06}  to fix
the final parameters. The newly developed IOPB-I set along with NL3
\cite{lala97}, FSUGarnet \cite{chai15} and G3 \cite{G3} are given for
comparison in Table \ref{table1}.  The calculated results of the binding
energy and charge radius are compared with the known experimental data
\cite{audi12,angeli13}.  It is to be noted that in the original
E-RMF parametrization, only five spherical nuclei were taken into
consideration while fitting the parameters with the binding energy,
charge radius and single particle energy \cite{furnstahl97}. However,
here, eight spherical nuclei are used for the fitting as listed in
Table \ref{table2}.

\begin{table}
\caption{The obtained new parameter set IOPB-I along with NL3 \cite{lala97},
FSUGarnet \cite{chai15} and G3 \cite{G3} sets are  listed. The nucleon mass 
$M$ is 939.0 MeV.  All the coupling constants are dimensionless, except 
$k_3$ which is in fm$^{-1}$.}
\scalebox{1.3}{
\begin{tabular}{cccccccccc}
\hline
\hline
\multicolumn{1}{c}{}
&\multicolumn{1}{c}{NL3}
&\multicolumn{1}{c}{FSUGarnet}
&\multicolumn{1}{c}{G3}
&\multicolumn{1}{c}{IOPB-I}\\
\hline
$m_{s}/M$  &  0.541  &  0.529&  0.559&0.533  \\
$m_{\omega}/M$  &  0.833  & 0.833 &  0.832&0.833  \\
$m_{\rho}/M$  &  0.812 & 0.812 &  0.820&0.812  \\
$m_{\delta}/M$   & 0.0  &  0.0&   1.043&0.0  \\
$g_{s}/4 \pi$  &  0.813  &  0.837 &  0.782 &0.827 \\
$g_{\omega}/4 \pi$  &  1.024  & 1.091 &  0.923&1.062 \\
$g_{\rho}/4 \pi$  &  0.712  & 1.105&  0.962 &0.885  \\
$g_{\delta}/4 \pi$  &  0.0  &  0.0&  0.160& 0.0 \\
$k_{3} $   &  1.465  & 1.368&    2.606 &1.496 \\
$k_{4}$  &  -5.688  &  -1.397& 1.694 &-2.932  \\
$\zeta_{0}$  &  0.0  &4.410&  1.010  &3.103  \\
$\eta_{1}$  &  0.0  & 0.0&  0.424 &0.0  \\
$\eta_{2}$  &  0.0  & 0.0&  0.114 &0.0  \\
$\eta_{\rho}$  &  0.0  & 0.0&  0.645& 0.0  \\
$\Lambda_{\omega}$  &  0.0  &0.043 &  0.038&0.024   \\
$\alpha_{1}$  &  0.0  & 0.0&   2.000&0.0  \\
$\alpha_{2}$  &  0.0  & 0.0&  -1.468&0.0  \\
$f_\omega/4$  &  0.0  & 0.0&  0.220&0.0 \\
$f_\rho/4$  &  0.0  & 0.0&    1.239&0.0 \\
$\beta_\sigma$  &  0.0  & 0.0& -0.087& 0.0  \\
$\beta_\omega$  &  0.0  & 0.0& -0.484& 0.0  \\
\hline
\hline
\end{tabular}}
\label{table1}
\end{table}

\begin{table}
\caption{The calculated binding energy per particle (B/A) and charge radius
	($R_c$) are compared with the available experimental data \cite{audi12,
	angeli13}. The predicted neutron-skin thickness 
	$\Delta r_{np}= R_n - R_p$ is also depicted with all the four models.}
\scalebox{1.0}{
\begin{tabular}{cccccccccc}
\hline
\hline
\multicolumn{1}{c}{Nucleus}&
\multicolumn{1}{c}{Obs.}&
\multicolumn{1}{c}{Expt.}&
\multicolumn{1}{c}{NL3}&
\multicolumn{1}{c}{FSUGarnet}&
\multicolumn{1}{c}{G3}&
\multicolumn{1}{c}{IOPB-I}\\
\hline
$^{16}$O&B/A & 7.976 &7.917&7.876 & 8.037&7.977  \\
         & R$_{c}$& 2.699 & 2.714&2.690   & 2.707&2.705  \\
         & R$_{n}$-R$_{p}$ & - & -0.026&-0.028  & -0.028&-0.027  \\
\\
$^{40}$Ca & B/A  & 8.551 & 8.540&8.528  &8.561&8.577  \\
         & R$_{c}$  & 3.478 &  3.466&3.438  & 3.459&3.458  \\
         & R$_{n}$-R$_{p}$  & - & -0.046&-0.051  & -0.049&-0.049  \\
\\
$^{48}$Ca & B/A  & 8.666 & 8.636&8.609  &8.671& 8.638 \\
         & R$_{c}$  & 3.477 & 3.443&3.426   & 3.466 &3.446 \\
         & R$_{n}$-R$_{p}$  & - &  0.229&0.169 & 0.174&0.202  \\
\\
$^{68}$Ni & B/A  & 8.682 & 8.698&8.692  & 8.690&8.707  \\
         & R$_{c}$  & - & 3.870 &3.861  &3.892&3.873  \\
         & R$_{n}$-R$_{p}$  & - &0.262 &0.184 &0.190&0.223  \\
\\
$^{90}$Zr & B/A  & 8.709 & 8.695&8.693  & 8.699&8.691  \\
         & R$_{c}$  & 4.269 & 4.253&4.231  & 4.276& 4.253 \\
         & R$_{n}$-R$_{p}$  & - & 0.115&0.065   & 0.068&0.091 \\
\\
$^{100}$Sn & B/A  & 8.258 & 8.301& 8.298   & 8.266&8.284 \\
         & R$_{c}$  & - & 4.469&4.426   &4.497&4.464  \\
         & R$_{n}$-R$_{p}$  & - & -0.073&-0.078  &  -0.079&-0.077 \\
\\
$^{132}$Sn & B/A  & 8.355 & 8.371&8.372  & 8.359&8.352 \\
         & R$_{c}$  & 4.709 & 4.697&4.687   & 4.732& 4.706 \\
         & R$_{n}$-R$_{p}$  & - & 0.349&0.224   & 0.243&0.287 \\
\\
$^{208}$Pb & B/A  & 7.867 &7.885& 7.902 & 7.863&7.870  \\
         & R$_{c}$  &5.501  & 5.509&5.496 & 5.541 &5.521 \\
         & R$_{n}$-R$_{p}$  & - &   0.283& 0.162 & 0.180 &0.221\\
\hline
\hline
\end{tabular}
\label{table2}}
\end{table}

\begin{table}
\caption{The nuclear matter properties such as binding energy per nucleon 
$\mathcal{E}_{0}$(MeV), saturation density $\rho_{0}$(fm$^{-3}$), 
incompressibility coefficient for symmetric nuclear matter $K$(MeV), 
effective mass ratio $M^*/M$, symmetry energy $J$(MeV) and linear density 
dependence of the symmetry energy $L$(MeV) at saturation.	}
\scalebox{1.3}{
\begin{tabular}{cccccccccc}
\hline
\hline
\multicolumn{1}{c}{}
&\multicolumn{1}{c}{NL3}
&\multicolumn{1}{c}{FSUGarnet}
&\multicolumn{1}{c}{G3}
&\multicolumn{1}{c}{IOPB-I}\\
\hline
	$\rho_{0}$ (fm$^{-3})$ &  0.148  &  0.153&  0.148&0.149  \\
$\mathcal{E}_{0}$(MeV)  &  -16.29  & -16.23 &  -16.02&-16.10  \\
$M^{*}/M$  &  0.595 & 0.578 &  0.699&0.593  \\
$J$(MeV)   & 37.43  &  30.95&   31.84&33.30  \\
$L$(MeV)  &  118.65  &  51.04 &  49.31&63.58 \\
$K_{sym}$(MeV)  &  101.34  & 59.36 & -106.07&-37.09 \\
$Q_{sym}$(MeV)  &  177.90  & 130.93&  915.47 &862.70  \\
$K$(MeV)  & 271.38  &  229.5&  243.96& 222.65 \\
$Q_{0} $(MeV)   &  211.94  & 15.76&   -466.61 &-101.37 \\
$K_{\tau}$(MeV)  &  -703.23  &  -250.41&-307.65 &-389.46  \\
$K_{asy}$(MeV)  & -610.56  & -246.89&  -401.97 &-418.58 \\
$K_{sat2}$(MeV)  & -703.23  &-250.41&  -307.65  &-389.46 \\
\hline
\hline
\end{tabular}
\label{table3}}
\end{table}

\begin{table*}
\hspace{0.1 cm}
	\caption{The binary neutron star masses ($m_1(M_\odot), m_2(M_\odot)$)
	and corresponding radii ($R_1$(km), $R_2$(km)), tidal love number 
	($(k_2)_1$, $(k_2)_2$), tidal deformabilities ($\lambda_1, \lambda_2$) 
	in $10^{36}$g cm$^2$ s$^2$ and 
	dimensionless tidal deformabilities ($\Lambda_1, \Lambda_2$). 
	$\tilde{\Lambda}$, $\delta\tilde\Lambda$, $\cal{M}$$_{c}(M_\odot)$ and 
	$\cal{R}$$_{c}(km)$ are the dimensionless tidal deformability,
	tidal correction, chirp mass and radius of the binary neutron star, 
	respectively.}
\renewcommand{\tabcolsep}{0.1 cm}
\renewcommand{\arraystretch}{1.2}
\begin{tabular}{ccccccccccccccccccccccccc}
\hline
\hline
\multicolumn{15}{ c }{}\\
EoS &       $m_1(M_\odot)$   &       $m_2(M_\odot)$ &$R_1$(km)  &       
	$R_2$(km)       &       $(k_2)_1$        &       $(k_2)_2$ &  $\lambda_1$ & 
	$\lambda_2$ &$\Lambda_1$&     $\Lambda_2$&$\tilde{\Lambda}$&$\delta\tilde\Lambda$&$\cal{M}$$_c(M_\odot)$&$\cal{R}$$_{c}(km)$\\
\hline
	&&&&&&&&&&&&&&\\
NL3 
&      1.20&      1.20&    14.702&    14.702&    0.1139&    0.1139&     7.826&     7.826&   2983.15&   2983.15&   2983.15&     0.000&      1.04&    10.350 \\
&      1.50&      1.20&    14.736&    14.702&    0.0991&    0.1139&     6.889&     7.826&    854.06&   2983.15&   1608.40&   220.223&      1.17&    10.214 \\
&      1.25&      1.25&    14.708&    14.708&    0.1118&    0.1118&     7.962&     7.962&   2388.82&   2388.82&   2388.82&     0.000&      1.09&    10.313 \\
&      1.30&      1.30&    14.714&    14.714&    0.1094&    0.1094&     7.546&     7.546&   1923.71&   1923.71&   1923.71&     0.000&      1.13&    10.271 \\
&      1.35&      1.35&    14.720&    14.720&    0.1070&    0.1070&     7.393&     7.393&   1556.84&   1556.84&   1556.84&     0.000&      1.18&    10.224 \\
&      1.35&      1.25&    14.720&    14.708&    0.1070&    0.1118&     7.393&     7.962&   1556.84&   2388.82&   1930.02&    91.752&      1.13&    10.268 \\
&      1.37&      1.25&    14.722&    14.708&    0.1061&    0.1118&     7.339&     7.962&   1452.81&   2388.82&   1863.78&   100.532&      1.14&    10.271 \\
&      1.40&      1.20&    14.726&    14.702&    0.1044&    0.1139&     7.231&     7.826&   1267.07&   2983.15&   1950.08&   183.662&      1.13&    10.262 \\
&      1.40&      1.40&    14.726&    14.726&    0.1044&    0.1044&     7.231&     7.231&   1267.07&   1267.07&   1267.07&     0.000&      1.22&    10.174 \\
&      1.42&      1.29&    14.728&    14.712&    0.1031&    0.1099&     7.147&     7.572&   1145.72&   1994.02&   1515.18&    95.968&      1.18&    10.192 \\
&      1.44&      1.39&    14.730&    14.724&    0.1027&    0.1049&     7.120&     7.259&   1108.00&   1311.00&   1204.83&    19.212&      1.23&    10.179 \\
&      1.45&      1.45&    14.732&    14.732&    0.1018&    0.1018&     7.064&     7.064&   1037.13&   1037.13&   1037.13&     0.000&      1.26&    10.124 \\
&      1.54&      1.26&    14.740&    14.708&    0.0969&    0.1114&     6.741&     7.668&    729.95&   2303.95&   1308.91&   168.202&      1.21&    10.179 \\
&      1.60&      1.60&    14.746&    14.746&    0.0937&    0.0937&     6.532&     6.532&    589.92&    589.92&    589.92&     0.000&      1.39&     9.979 \\	
	&&&&&&&&&&&&&&\\
FSUGarnet 
&      1.20&      1.20&    12.944&    12.944&    0.1090&    0.1090&     3.961&     3.961&   1469.32&   1469.32&   1469.32&     0.000&      1.04&     8.983 \\
&      1.50&      1.20&    12.972&    12.944&    0.0893&    0.1090&     3.282&     3.961&    408.91&   1469.32&    784.09&   111.643&      1.17&     8.847 \\
&      1.25&      1.25&    12.958&    12.958&    0.1062&    0.1062&     3.880&     3.880&   1193.78&   1193.78&   1193.78&     0.000&      1.09&     8.977 \\
&      1.30&      1.30&    12.968&    12.968&    0.1030&    0.1030&     3.777&     3.777&    945.29&    945.29&    945.29&     0.000&      1.13&     8.910 \\
&      1.35&      1.35&    12.974&    12.974&    0.0998&    0.0998&     3.666&     3.666&    761.13&    761.13&    761.13&     0.000&      1.18&     8.860 \\
&      1.35&      1.25&    12.974&    12.958&    0.0998&    0.1062&     3.666&     3.880&    761.13&   1193.78&    955.00&    49.744&      1.13&     8.920 \\
&      1.37&      1.25&    12.976&    12.958&    0.0986&    0.1062&     3.629&     3.880&    710.62&   1193.78&    922.54&    53.853&      1.14&     8.924 \\
&      1.40&      1.20&    12.978&    12.944&    0.0965&    0.1090&     3.552&     3.961&    622.06&   1469.32&    959.22&    90.970&      1.13&     8.904 \\
&      1.40&      1.40&    12.978&    12.978&    0.0965&    0.0965&     3.552&     3.552&    622.06&    622.06&    622.06&     0.000&      1.22&     8.825 \\
&      1.42&      1.29&    12.978&    12.966&    0.0949&    0.1038&     3.495&     3.803&    565.47&   1001.18&    755.10&    50.492&      1.18&     8.867 \\
&      1.44&      1.39&    12.978&    12.978&    0.0939&    0.0973&     3.456&     3.582&    531.54&    653.60&    589.65&    14.148&      1.23&     8.823 \\
&      1.45&      1.45&    12.978&    12.978&    0.0931&    0.0931&     3.427&     3.427&    507.70&    507.70&    507.70&     0.000&      1.26&     8.776 \\
&      1.54&      1.26&    12.964&    12.960&    0.0862&    0.1057&     3.157&     3.864&    343.73&   1146.73&    638.35&    88.892&      1.21&     8.817 \\
&      1.60&      1.60&    12.944&    12.944&    0.0816&    0.0816&     2.964&     2.964&    266.20&    266.20&    266.20&     0.000&      1.39&     8.511 \\	
	&&&&&&&&&&&&&&\\
G3 
&      1.20&      1.20&    12.466&    12.466&    0.1034&    0.1034&     3.114&     3.114&   1776.65&   1776.65&   1776.65&     0.000&      1.04&     9.331 \\
&      1.50&      1.20&   112.360&    12.466&    0.0800&    0.1034&     2.309&     3.114&    284.92&   1776.65&    803.43&   191.605&      1.17&     8.890 \\
&      1.25&      1.25&    12.460&    12.460&    0.1001&    0.1001&     3.007&     3.007&    939.79&    939.79&    939.79&     0.000&      1.09&     8.557 \\
&      1.30&      1.30&    12.448&    12.448&    0.0962&    0.0962&     2.875&     2.875&    728.07&    728.07&    728.07&     0.000&      1.13&     8.457 \\
&      1.35&      1.35&    12.434&    12.434&    0.0925&    0.0925&     2.750&     2.750&    582.26&    582.26&    582.26&     0.000&      1.18&     8.398 \\
&      1.35&      1.25&    12.434&    12.460&    0.0925&    0.1001&     2.750&     3.007&    582.26&    939.79&    742.29&    43.064&      1.13&     8.482 \\
&      1.37&      1.25&    12.428&    12.460&    0.0909&    0.1001&     2.696&     3.007&    530.66&    939.79&    709.72&    49.144&      1.14&     8.468 \\
&      1.40&      1.20&    12.416&    12.466&    0.0859&    0.1034&     2.613&     3.114&    461.03&   1776.65&    976.80&   183.274&      1.13&     8.937 \\
&      1.40&      1.40&    12.416&    12.416&    0.0859&    0.0859&     2.613&     2.613&    461.03&    461.03&    461.03&     0.000&      1.22&     8.312 \\
&      1.42&      1.29&    12.408&    12.450&    0.0868&    0.0972&     2.553&     2.905&    417.96&    772.17&    571.87&    43.226&      1.18&     8.387 \\
&      1.44&      1.39&    12.398&    12.420&    0.0854&    0.0894&     2.501&     2.643&    384.42&    484.90&    432.22&    12.671&      1.23&     8.292 \\
&      1.45&      1.45&    12.932&    12.392&    0.0846&    0.0846&     2.472&     2.472&    367.04&    367.04&    367.04&     0.000&      1.26&     8.225 \\
&      1.54&      1.26&    12.334&    12.458&    0.0769&    0.0992&     2.194&     2.976&    239.49&    883.46&    474.83&    75.175&      1.21&     8.311 \\
&      1.60&      1.60&    12.280&    12.280&    0.0716&    0.0716&     2.000&     2.000&    179.63&    179.63&    179.63&     0.000&      1.39&     7.867 \\	
	&&&&&&&&&&&&&&\\

\hline
\hline
\end{tabular}
\label{table4}
\end{table*}

\begin{table*}
\hspace{0.1 cm}
        \caption{Table \ref{table4} is continued \dots}
\renewcommand{\tabcolsep}{0.1 cm}
\renewcommand{\arraystretch}{1.2}
\begin{tabular}{ccccccccccccccccccccccccc}
\hline
\hline
\multicolumn{15}{ c }{}\\
EoS &       $m_1(M_\odot)$   &       $m_2(M_\odot)$ &$R_1$(km)  &
        $R_2$(km)       &       $(k_2)_1$        &       $(k_2)_2$ &  $\lambda_1$ &
        $\lambda_2$ &$\Lambda_1$&     $\Lambda_2$&$\tilde{\Lambda}$&$\delta\tilde\Lambda$&$\cal{M}$$_c(M_\odot)$&$\cal{R}$$_{c}(km)$\\
\hline
        &&&&&&&&&&&&&&\\
IOPB-I
&      1.20&      1.20&    13.222&    13.222&    0.1081&    0.1081&     4.369&     4.369&   1654.23&   1654.23&   1654.23&     0.000&      1.04&     9.199 \\
&      1.50&      1.20&    13.236&    13.222&    0.0894&    0.1081&     3.631&     4.369&    449.62&   1654.23&    875.35&   128.596&      1.17&     9.044 \\
&      1.25&      1.25&    13.230&    13.230&    0.1053&    0.1053&     4.268&     4.268&   1310.64&   1310.64&   1310.64&     0.000&      1.09&     9.146 \\
&      1.30&      1.30&    13.238&    13.238&    0.1024&    0.1024&     4.162&     4.162&   1053.07&   1053.07&   1053.07&     0.000&      1.13&     9.105 \\
&      1.35&      1.35&    13.240&    13.240&    0.0995&    0.0995&     4.050&     4.050&    857.53&    857.53&    857.53&     0.000&      1.18&     9.074 \\
&      1.35&      1.25&    13.240&    13.230&    0.0995&    0.1053&     4.050&     4.268&    857.53&   1310.64&   1060.81&    49.565&      1.13&     9.110 \\
&      1.37&      1.25&    13.242&    13.230&    0.0938&    0.1053&     4.004&     4.268&    791.92&   1310.64&   1019.60&    56.371&      1.14&     9.104 \\
&      1.40&      1.20&    13.242&    13.222&    0.0960&    0.1081&     3.911&     4.369&    680.79&   1654.23&   1067.64&   107.340&      1.13&     9.097 \\
&      1.40&      1.40&    13.242&    13.242&    0.0960&    0.0960&     3.911&     3.911&    680.79&    680.79&    680.79&     0.000&      1.22&     8.986 \\
&      1.42&      1.29&    13.242&    13.236&    0.0949&    0.1030&     3.864&     4.184&    632.31&   1099.78&    835.91&    52.836&      1.18&     9.049 \\
&      1.44&      1.39&    13.242&    13.242&    0.0935&    0.0969&     3.806&     3.946&    578.47&    719.80&    645.73&    17.094&      1.23&     8.985 \\
&      1.45&      1.45&    13.240&    13.240&    0.0927&    0.0927&     3.771&     3.771&    549.06&    549.06&    549.06&     0.000&      1.26&     8.915 \\
&      1.54&      1.26&    13.230&    13.232&    0.0868&    0.1047&     3.516&     4.247&    384.65&   1253.00&    703.58&    94.735&      1.21&     8.991 \\
&      1.60&      1.60&    13.212&    12.212&    0.0823&    0.0823&     3.314&     3.314&    296.81&    296.81&    296.81&     0.000&      1.39&     8.698 \\
\hline
\hline
\end{tabular}
\label{table5} 
\end{table*}

\begin{figure}[!b]
        \includegraphics[width=1.1\columnwidth]{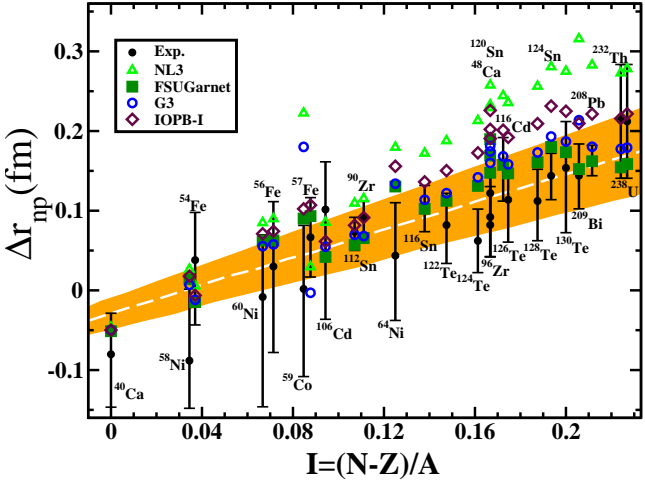}
        \caption{(color online) The neutron-skin thickness as a function
of the asymmetry parameter. Results obtained with the parameter set IOPB-I are compared with those of the set NL3 \cite{lala97}, FSUGarnet \cite{chai15}, G3 
\cite{G3} and experimental values \cite{thick}. The shaded region is calculated
using Eq.\ref{eq:rnp}.}
        \label{thick}
\end{figure}


\begin{figure}[!b]
	\includegraphics[width=1.\columnwidth]{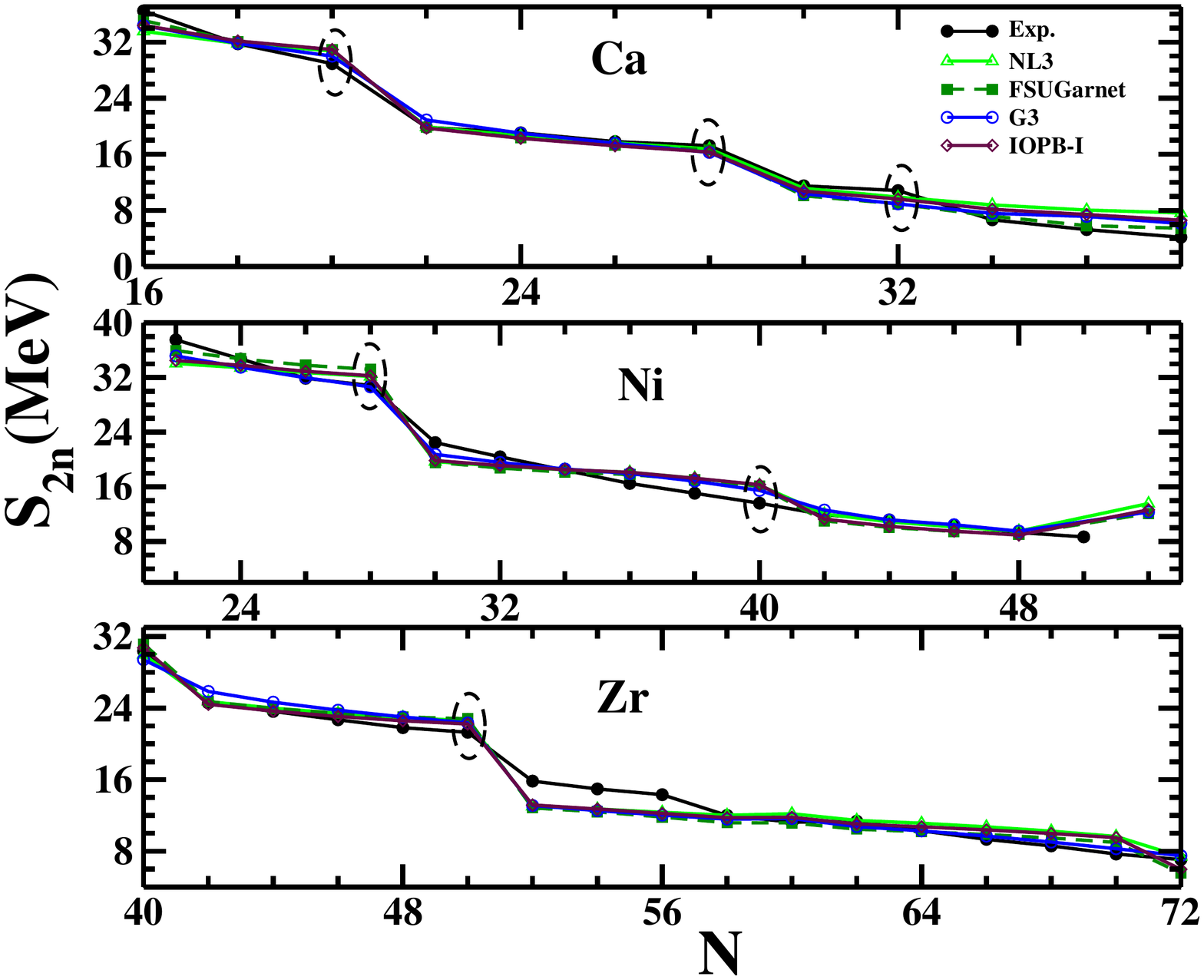}
	\includegraphics[width=1.\columnwidth]{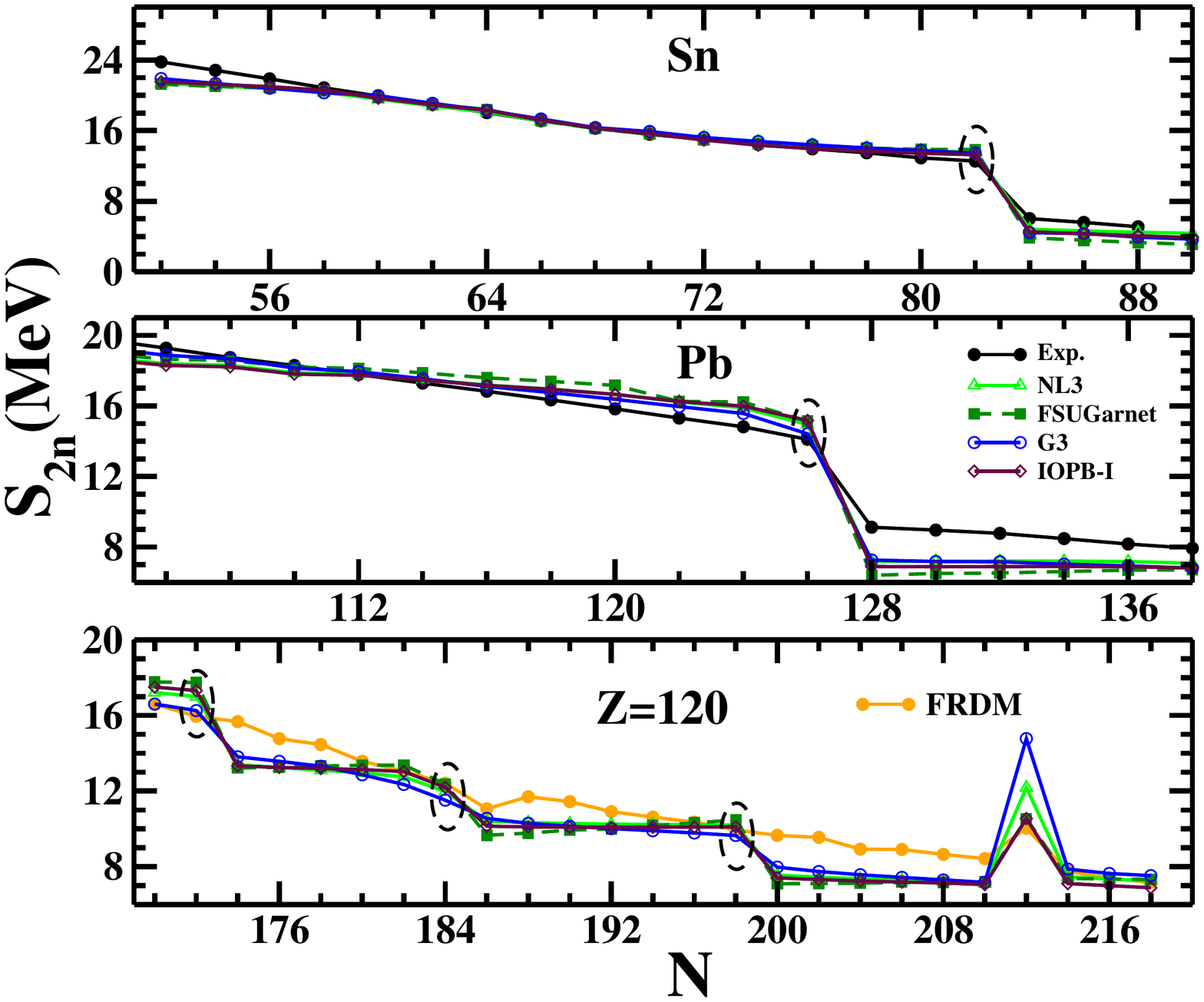}
	\caption{(color online) The two-neutron separation energy as a 
function of neutron number for the isotopic series of  Ca, Ni, Zr, Sn and Pb 
nuclei with NL3 \cite{lala97}, FSUGarnet \cite{chai15}, G3 \cite{G3}, 
FRDM \cite{moller} and experimental data \cite{audi12} whenever available. The 
dotted circle represents the magicity of the nuclei.}
	\label{s2n}
\end{figure}

\section{RESULTS AND DISCUSSIONS }\label{result}

In this section we discuss our calculated results for finite
nuclei, infinite nuclear matter and neutron stars. For finite nuclei,
binding energy, rms radii for neutron and proton distributions, two
neutron separation energy and neutron-skin thickness are analyzed.
Similarly, for infinite nuclear matter system, the binding energy
per particle for symmetric and asymmetric nuclear matter including
pure neutron matter at both sub-saturation and supra-saturation densities are
compared with other theoretical results and experimental data. The
parameter set IOPB-I is also applied to study the structure of
neutron star using $\beta-$equilibrium and charge neutrality conditions.

\subsection{FINITE NUCLEI}

\begin{center}
\bf (i) Binding energies, charge radii and neutron-skin thickness
\end{center}
We have used eight spherical nuclei to fit the experimental ground-state
binding energies and charge radii using SAM. The calculated results are
listed in Table \ref{table2} and compared with other theoretical models
as well as experimental data \cite{audi12,angeli13}. It can be seen
that the NL3 \cite{lala97}, FSUGarnet \cite{chai15} and G3 \cite{G3}
successfully reproduce the energies and charge radii as well. Although,
the {\it "mean-field models are not expected to work well for the
light nuclei " but the results deviate only marginally for the
ground state properties  for light nuclei} \cite{patra93}. We noticed
that both the binding energy and charge radius of $^{16}$O are well
produced by IOPB-I. However, the charge radii of $^{40,48}$Ca slightly
underestimate the data. We would like to emphasize that it is an open
problem to mean-field models to predict the evolution of charge radii
of $^{38-52}$Ca (see Fig. 3 in Ref. \cite{gar16}).

The excess of neutrons give rise to a neutron-skin thickness.
The neutron-skin thickness $\Delta r_{np}$ is defined as
\begin{equation} \Delta r_{np}=\langle r^2 \rangle^{1/2}_{n} -
\langle r^2 \rangle^{1/2}_{p} =R_n-R_p .
 \end{equation} 
with $R_n$ and $R_p$ being the rms radii for the neutron and proton
distributions, respectively. 
The $\Delta r_{np}$, strongly correlated with the slope of the symmetry
energy \cite{pg10,brown00,maza11}, can probe isovector  part of the
nuclear interaction. However, there is a large uncertainty in the
experimental measurement of neutron distribution radius of the finite
nuclei.  The current values of neutron radius and neutron-skin thickness
of $^{208}$Pb are 5.78$^{+0.16}_{-0.18}$ and 0.33$^{+0.16}_{-0.18}$
fm, respectively \cite{abra12}. This error bar is too large to provide significant
constraints on the density-dependent of symmetry energy. It is expected
that PREX-II result will give us the neutron radius of $^{208}$Pb within
$1\%$ accuracy. The inclusion of some isovector dependent terms in the
Lagrangian density is needed which would provide the freedom to refit
the coupling constants within the experimental data without compromising
the quality of fit.  The addition of $\omega-\rho$ cross-coupling into
the Lagrangian density controls the neutron-skin thickness of $^{208}$Pb
as well as for other nuclei.  In Fig. \ref{thick}, we show neutron-skin
thickness $\Delta r_{np}$ for $^{40}$Ca to $^{238}$U nuclei as a function
of proton-neutron asymmetry $I=(N-Z)/A$. The calculated results of $\Delta
r_{np}$ for NL3, FSUGarnet, G3 and IOPB-I parameter sets are compared with
the corresponding experimental data \cite{thick}.  Experiments have been
done with antiprotons at CERN and  the $\Delta r_{np}$ are extracted 
for 26 stable nuclei ranging from $^{40}$Ca to $^{238}$U  as displayed in
the figure along with the error bars. The trend of the data points show
approximately linear dependence of neutron-skin thickness on the relative
neutron excess $I$ of nucleus that can be fitted by \cite{thick,xavier14}:

\begin{eqnarray}
\Delta r_{np} = (0.90\pm0.15)I+(-0.03\pm0.02) \; {\text fm}.
\label{eq:rnp}
\end{eqnarray}

The values of  $\Delta r_{np}$ obtained with IOPB-I  for some of the
nuclei slightly deviate from the shaded region as can be seen from Fig.
\ref{thick}.  This is because, IOPB-I has a smaller strength of
$\omega-\rho$ cross-coupling as compared to FSUGarnet. Recently,
F. Fattoyev {\it et. al.} constraint the upper limit of $\Delta
r_{np}\lesssim0.25$ fm for $^{208}$Pb nucleus with the help of GW170817
observation data \cite{fatt17}.  The calculated values of neutron-skin
thickness for the $^{208}$Pb nucleus are 0.283, 0.162, 0.180 and 0.221
fm for the NL3, FSUGarnet, G3 and IOPB-I parameter sets respectively. The
proton elastic scattering experiment  has recently measured neutron-skin
thickness $\Delta r_{np}=0.211^{+0.054}_{-0.063} $ fm for $^{208}$Pb
\cite{zeni10}.  Thus values of $\Delta r_{np} = 0.221$ for IOPB-I  are
consistent with recent prediction of neutron-skin thickness.

\begin{center}
\bf (ii)  Two-neutron separation energy $S_{2n}(Z,N)$
\end{center}
The large shell gap in single particle energy levels is an indication of
the magic number.  This is responsible for the extra stability for the
magic nuclei.  The extra stability for a particular  nucleon number
can be understood  from the sudden fall in the two-neutron separation
energy $S_{2n}$. The $S_{2n}$ can be estimated  by the difference in
ground state binding energies of two isotopes {\it i.e.},

\begin{equation}
S_{2n}(Z,N) = BE(Z,N) - BE(Z,N-2).
\end{equation}

In Fig. \ref{s2n}, we display results for the $S_{2n}$ as a function of
neutron numbers for  Ca, Ni, Zr, Sn, Pb and Z=120 isotopic chains. The
calculated results are compared with the finite range droplet model (FRDM)
\cite{moller} and latest experimental data \cite{audi12}. From the figure,
it is clear that there is an evolution of magicity as one moves from the valley
of stability to the {\it drip-line}. In all cases, the $S_{2n}$ values
decrease gradually with increase in neutron number.  The experimental
manifestation of large shell gaps at  neutron number N = 20, 28(Ca),
28(Ni), 50(Zr), 82(Sn) and 126(Pb) are reasonably well reproduced by the
four relativistic sets. 
Figure \ref{s2n} shows that the experimental 
$S_{2n}$ of $^{50-52}$Ca are in good agreement with the prediction of NL3 set. 
It is interesting to note that all sets predict the sub-shell closure
at N=40 for Ni isotopes.  Furthermore,  the two-neutron separation
energy for the isotopic chain of nuclei with  Z=120 is also displayed
in Fig. \ref{s2n}. For the isotopic chain of Z=120, no
experimental information exit. The only comparison can be made with
theoretical models such as the FRDM \cite{audi12}. At N=172, 184 and
198 sharp falls in separation energy is seen for all forces,
which have been predicted by various theoretical models in the superheavy
mass region \cite{rutz,gupta,patra99,mehta}. It is to be noted that
the isotopes with Z=120 are shown to be spherical in their ground state
\cite{mehta}. 
In a detail calculation, Bhuyan and Patra using both RMF and Skyrme-Hartree-Fock formalisms, predicted that Z=120 could be the next magic number after Z=82 in 
the superheavy region \cite{bunu12}. Thus, the deformation effects may not 
affect the results for Z=120. Therefore, a future mass measurement of 
$^{292,304,318}$120 would confirm a key test for the theory,
as well as direct information about the closed-shell behavior at N=172,
184 and 198.

\begin{figure}
\includegraphics[width=1\columnwidth]{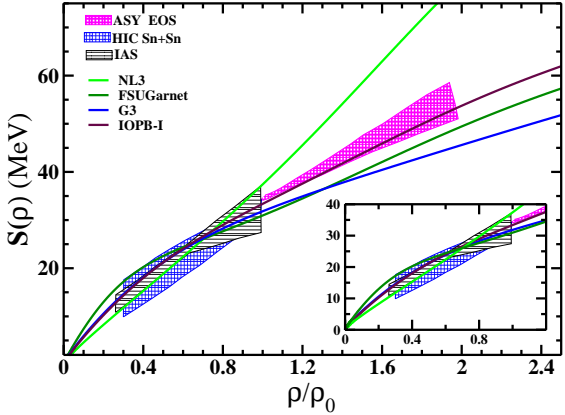}
\caption{(color online) Density dependent symmetry energy from Eq. (\ref{eq30}) 
with different E-RMF parameter sets along with IOPB-I parametrization. The 
shaded region is the symmetry energy from IAS \cite{ias}, HIC Sn+Sn \cite{snsn} and ASY-EoS experimental data \cite{asym}. The zoomed pattern of the symmetry energy at low densities is shown in the inset.}
        \label{sym}
\end{figure}

\subsection{Infinite Nuclear  Matter}
The nuclear incompressibility $K$  
determines the extent to which the nuclear matter can be compressed. 
This plays  important role in the nuclear equation of state (EoS). 
Currently, the accepted value of $K=240\pm20$ MeV has been determined from
isoscalar giant monopole resonance (ISGMR) for $^{90}$Zr and $^{208}$Pb nuclei 
\cite{colo14,piek14}. 
For our parameter set IOPB-I, we get $K=222.65$ MeV.
The density-dependent symmetry energy $S(\rho)$ is 
determined from Eq.(\ref{eq30}) using IOPB-I along with three adopted models. 
The calculated results of the symmetry energy coefficient ($J$), the slope of 
symmetry energy ($L$) and other saturation properties are listed in 
Table \ref{table3}. 
We find that in case of IOPB-I, $J=33.3$ MeV and $L=63.6$ MeV. These values 
are compatible with $J=31.6\pm2.66$ MeV and $L=58.9\pm16$ MeV obtained by 
various terrestrial experimental informations and astrophysical observations 
\cite{li13}.

Another important constraint $K_\tau$ has been suggested which lies in the
range of -840 MeV to -350 MeV \cite{stone14,pear10,li10} by various 
experimental data on isoscalar giant monopole resonance, which we can calculate 
from Eq.(\ref{eq34}). It is to be noticed that the calculated values of 
$K_\tau$ are -703.23, -250.41, -307.65 and -389.46 MeV for NL3, FSUGarnet, 
G3 and IOPB-I parameter sets, respectively.  
The ISGMR measurement has been
investigated in a series of $^{112-124}$Sn isotopes, which  extracted
the value of $K_\tau=-395\pm40$ MeV \cite{garg07}. It is found that
$K_\tau=-389.46$ MeV for IOPB-I set is consistent with GMR measurement \cite{garg07}. 
In the absence of cross-coupling, $S(\rho)$ of NL3 is stiffer at low
and high-density regime as displayed in Fig. \ref{sym}.  Alternatively,
presence of cross-coupling of $\rho$ mesons to the $\omega$ (in case
of FSUGarnet and IOPB-I) and $\sigma$ mesons (in case of G3) yields the
softer symmetry energy at low-density which are consistent with HIC Sn+Sn
\cite{snsn}, IAS \cite{ias} data as shown in the figure. However,
IOPB-I has softer $S(\rho)$ in comparison to NL3 parameter set at
higher density which lies inside the shaded region of ASY-EoS experimental
data \cite{asym} .

\begin{figure}[!t]
\includegraphics[width=1\columnwidth]{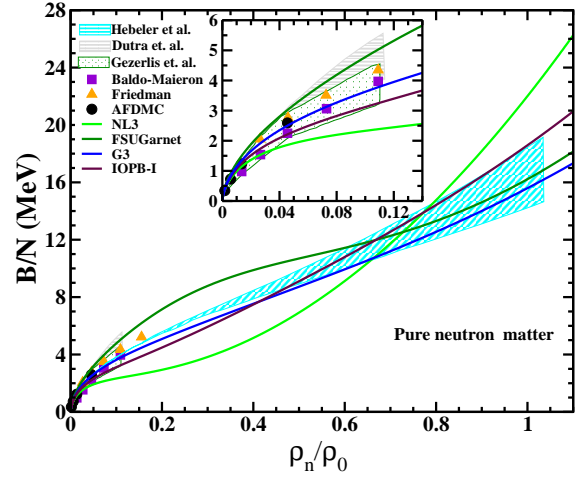}
\caption{(color online) The energy per neutron as a function of
neutron density with  NL3 \cite{lala97}, FSUGarnet \cite{chai15},G3 
\cite{G3} and IOPB-I parameter sets. Other curves and shaded region represents 
the results for various microscopic approaches such as Baldo-Maieron 
\cite{baldo}, Friedman \cite{friedman}, Auxiliary-field diffusion Monte 
Carlo \cite{gando}, Dutra \cite{dutra12}, Gezerlis \cite{geze10} and 
Hebeler \cite{hebe13}.}
	\label{sat}
\end{figure}

 Next, we display in Fig. \ref{sat} the binding energy per neutron (B/N)
as a function of the neutron density. Here, special attention is needed
to build nucleon-nucleon interaction to fit the data at sub-saturation
density.  For example, the EoS of pure neutron matter (PNM) at low
density obtained within the variational method, which is obtained
with a Urbana $v14$ interaction \cite{friedman}. In this regard,
the effective mean field models also fulfill this demand to some
extent \cite{chai15,mondal16,G3}. The cross-coupling $\omega-\rho$
plays an important role at low-density of the PNM.  The low-density
(zoomed pattern) nature of the FSUGarnet, G3 and IOPB-I are in
harmony with the results obtained by microscopic calculations
\cite{baldo,friedman,gando,dutra12,geze10}, while the results for
NL3 deviate from the shaded region at low as well as high density
regions. We also find a very good agreement for FSUGarnet, G3 and IOPB-I
at higher densities, which have been obtained with chiral NN and 3N
interactions \cite{hebe13}.

\begin{figure}
\includegraphics[width=1\columnwidth]{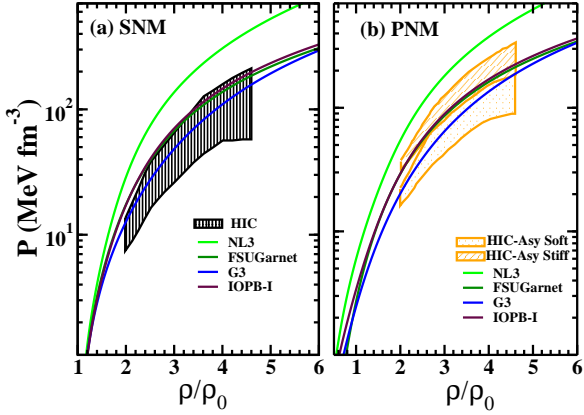}
\caption{(color online) Pressure as a function of baryon density  
for IOPB-I force. The results with NL3 \cite{lala97}, FSUGarnet \cite{chai15}, 
G3 \cite{G3} are compared with the EoS extracted from the analysis 
\cite{daniel02} for the (a) symmetric nuclear matter (SNM) and (b) pure neutron matter (PNM). }
	\label{eos}
\end{figure}

In Fig. \ref{eos}, we show the calculated pressure $P$ for the symmetric
nuclear matter (SNM) and PNM with the baryon density for the four
E-RMF models and then  are compared with the experimental flow data
\cite{daniel02}.  It is seen from Fig. \ref{eos}(a) that the SNM EoS for G3 
parameter set is in excellent agreement with the flow data for the entire 
density range. The SNM EoSs for the  FSUGarnet and IOPB-I are also compatible
 with the experimental HIC data  but they are stiffer relative to the EoS for 
the G3 parametarization.
In Fig. \ref{eos}(b), the bounds on the PNM equation of states
(EoSs) are divided into two categories (i) the upper one corresponds to
a strong density dependence of symmetry energy $S(\rho)$ (HIC-Asy Stiff)
and (ii) the lower one corresponds to the weakest $S(\rho)$(HIC-Asy
Soft) \cite{daniel02,prak88}.  Our parameter set IOPB-I along with G3
and FSUGarnet are reasonably in good agreement with experimental flow
data. The PNM EoS for the IOPB-I model is quite stiffer than  G3 at
high densities.

\begin{center}
        {\bf C. \; Neutron stars}
\end{center}
\begin{center}
	{\bf (i) \; Predicted equation of states}
\end{center}

\begin{figure}
\includegraphics[width=1\columnwidth]{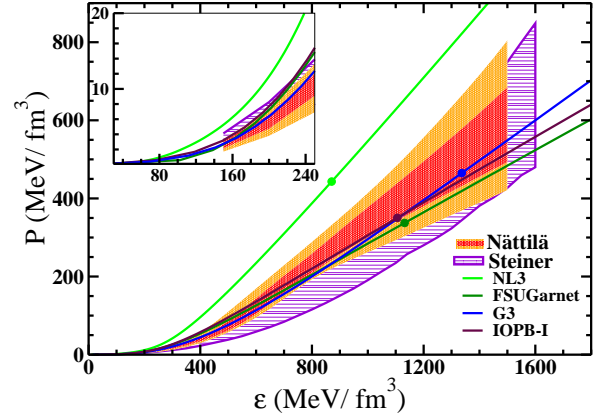}
\caption{(color online) The equations of states with NL3, FSUGarnet, G3 and 
IOPB-I for nuclear matter under charge neutrality as well as the 
$\beta$-equilibrium condition. The shaded region (violet) represents the 
observational constraint at $r_{ph} = R$ with uncertainty of 2$\sigma$ \cite{steiner10}. Here, $R$ and $r_{ph}$ are the neutron star radius and the 
photospheric radius, respectively. Other shaded region (red+orange) represents the QMC+Model A equation of state of cold dense matter with $95\%$ confidence limit 
\cite{joonas}. The region zoomed near the origin is shown in the inset.}
        \label{ep2}
\end{figure}

We have solved Eqs.(\ref{eq20}) and (\ref{eq21}) for the energy 
density and pressure of the {\it beta-equilibrated} charge neutral neutron star matter.
Fig. \ref{ep2} displays the pressure as a function of energy
density for the IOPB-I along with NL3, FSUGarnet and G3 sets.
The solid circles are the central pressure and energy
density corresponding to the maximum mass of the neutron star obtained
from the above equations of state. The shaded region of the EoS can be
divided into two parts as follows:

(i) N\"attli\"a {\it et. al.} applied the Bayesian cooling tail method to 
constraint (1$\sigma$ and 2$\sigma$ confidenced limit) the EoS of cold densed 
matter inside the neutron stars \cite{joonas}.\\
(ii) Steiner {\it et. al.} have determined an empirical densed matter 
EoS with $95\%$ confidence limit from a heterogeneous data set containing PRE 
bursts and quiescent thermal emission from X-ray transients \cite{steiner10}.

From Fig. \ref{ep2}, it is clear that IOPB-I and FSUGarnet EoSs are
similar at high-density but they differ remarkably at low
densities as shown in the zoomed area of the inset.
The  NL3 set yields the stiffer EoS. Moreover, IOPB-I shows
the stiffest EoS up to energy densities $\mathcal{E}$$\lesssim$700
MeV fm$^{-3}$. It can be seen that the results of IOPB-I at very
high densities $\mathcal{E} \sim 400-1600$ MeV fm$^{-3}$ consistent
with the EoS obtained by N\"attil\"a and Steiner {\it et. al.}
\cite{joonas,steiner10}. But, FSUGarnet has a softer EoS at low energy
densities $\mathcal{E}$$\lesssim$200 MeV fm$^{-3}$ and stiffer EoS
at intermediate energy densities as compared to that for the G3 set. 
One can conclude from  Table \ref{table3} that the symmetry energy elements
$L$ and $K_{sym}$ are smaller in G3 model compare to IOPB-I, FSUGarnet, and
NL3. Hence, It yields the symmetry energy softer at higher
density.

\begin{center}
\bf (ii) Mass-radius and tidal deformability of neutron star 
\end{center}
After fixing the equation of state for the various parameter sets, we 
extend our study to calculate the mass, radius and tidal deformability of a 
non-rotating neutron star.  Placing a spherical star in a static external
quadrupolar tidal field $\mathcal{E}_{ij}$ results in deformation of the star
along with  quadrupole deformation which is the leading order perturbation.
Such a deformation is measured by \cite{tanja1} 
\begin{eqnarray}
	\lambda = -\frac {Q_{ij}}{\mathcal{E}_{ij}}=\frac{2}{3} k_2 R^{5} \; ;
	\label{eq40}
\end{eqnarray}
\begin{eqnarray}
	\Lambda=\frac{2k_2}{3C^5}.
	\label{eq41}
\end{eqnarray}
 Where $Q_{ij}$ is the induced quadrupole moment of a star in binary, and
 $\cal{E}$$_{ij}$ is the static external quadrupole tidal field of the companion
 star. $\lambda$ is the tidal deformability parameter depending on the EoS via 
both the neutron star (NS) radius and a dimensionless quantity $k_2$, called 
the second Love number \cite{flan,tanja1}. $\Lambda$ is the dimensionless 
version of $\lambda$, and $C$ is the compactness parameter ($C=M/R$). However, 
in general relativity (GR) we have to distinguish $k_2$ between gravitational fields generated by masses (electric-type) and those generated by the motion of masses, {\it i.e.} mass currents (magnetic-type) \cite{phil,bharat}. The electric tidal Love is found from the following 
expression \cite{tanja1}:
\begin{widetext}
\begin{eqnarray}
k_{2} = \frac{8}{5}(1-2C)^{2}C^{5}[2C(y-1)-y+2]
\Big\{ 2C(4(y+1)C^{4}+(6y-4)C^{3}+
\nonumber \\
(26-22y)C^{2}+3(5y-8)C-3y+6)-3(1-2C)^{2}(2C(y-1)-y+2)log\Big( \frac{1}{1-2C}\Big)\Big\}^{-1} \; .
	\label{eq42}
\end{eqnarray}
\end{widetext}

\begin{figure}
\includegraphics[width=1.1\columnwidth]{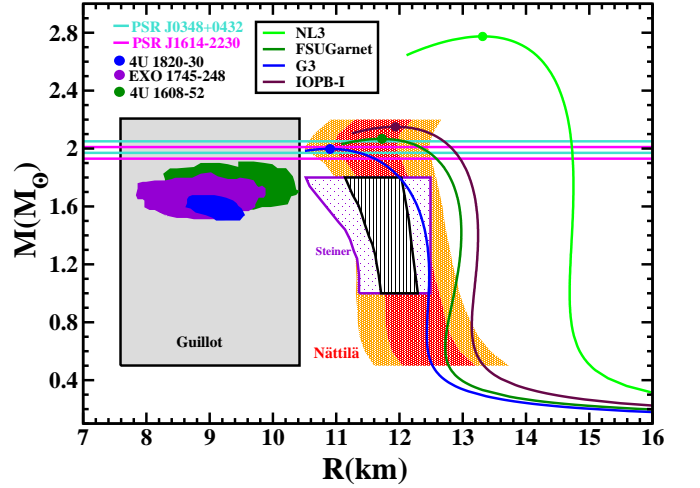}
\caption{(color online) The mass-radius profile predicted by NL3, FSUGarnet, G3 and IOPB-I. The recent observational constraints on neutron-star masses \cite{demo10,antoni13} and radii \cite{ozel10,steiner10,sule11,joonas} are also shown. }
	\label{MR}
\end{figure}

The value of $y\equiv y(R)$ can be computed by solving the following first 
order differential equation \cite{tanja,bharat}:
\begin{eqnarray}
r\frac{d y(r)}{d r} + y(r)^{2} + y(r) F(r) + r^{2} Q(r)=0,
	\label{eq43}
\end{eqnarray}
with,
\begin{eqnarray}
 F(r)=\frac{r - 4 \pi r^{3} [{\mathcal{E}}(r)-P(r)]}{r-2 M(r)},\\
 Q(r)=\frac{4 \pi r(5{\mathcal{E}}(r)+9 P(r)+\frac{{\mathcal{E}}(r)+P(r)}
{{\partial P(r)}/ {\partial{\mathcal{E}}(r)}}-\frac{6}{4 \pi r^{2}})}{r-2 M(r)}
\nonumber \\
-4 \Big[\frac {M(r)+4 \pi r^{3} P(r)}{r^{2}(1- 2 M(r)/r)}\Big]^{2}.
\end{eqnarray}
To estimate the tidal deformability $\lambda$ of a single star, Eq.(\ref{eq43}) must be integrated simultaneously with Tolman-Oppenheimer-Volkov (TOV) equations 
\cite{tov},  {\it i.e.}

\begin{eqnarray}
\frac{d P(r)}{d r}=-\frac{[{\mathcal{E}}(r)+P(r)][M(r)+{4\pi r^3 P(r)}]}{r^2(1-\frac{2M(r)}{ r})}, 
	\label{eq46}
\end{eqnarray}
and
\begin{eqnarray}
\frac{d M(r)}{d r}={4\pi r^2 {\mathcal{E}}(r)}.
	\label{eq47}
\end{eqnarray}
For a given EoS and from the boundary conditions $P(0) = P_{c}$, $M(0)$=0, and
$y(0)=2$, where $P_{c}$, M(0), and  y(0) are the central pressure, mass and 
dimensionless quantity.  To obtain the tidal Love number, we solve this set  
of Eqs. (\ref{eq40}-\ref{eq47}) for a given EoS of the
star at $r$=0. The value of $r$(= $R$), where the pressure vanishes
defines the surface of the star.  Thus, at each central density we can
uniquely determine a mass $M$, a radius R  and tidal Love number $k_2$
of the isolated neutron star using the chosen EoSs. In Fig. \ref{MR},
the horizontal bars in cyan and magenta colours include the results
from the precisely measured neutron stars masses, such as PSR J1614-2230
with mass $M=1.97\pm 0.04M_\odot$ \cite{demo10} and  PSR J0348+0432 with
$M=2.01\pm 0.04 M_\odot$ \cite{antoni13}.  These observations imply  that
the  maximum mass predicted by any theoretical model should reach the
limit $\sim 2.0 M_\odot$.  We also depict the $1\sigma$ and $2\sigma$
empirical mass-radius constraints for the cold densed matter inside the
NS which have been obtained from a Bayesian analysis of type-I X-ray burst
observation \cite{joonas}. Similar approach is applied by Steiner {\it et.
al.}, but they obtained the mass-radius from six sources, {\it i.e.},
three from transient low-mass X-ray binaries and three from type-I X-ray
bursters with photospheric radius \cite{steiner10}.

The NL3 model of RMF theory suggests a larger and massive NS with mass
$2.77 M_\odot$ and the corresponding NS radius to be 13.314 km, which is
larger than the best observational radius estimates \cite{steiner10,joonas}. 
Hence it is clear that the new RMF have been developed either through 
density-dependent couplings \cite{roca11} or higher order couplings 
\cite{G3,chai15}.  These models successfully reproduce the ground-state 
properties of finite nuclei, nuclear matter saturation properties and also for 
the maximum mass of the neutron stars. Another important advantage of these 
models is that they are consistent with the sub-saturation density of
the pure neutron matter.  L. Rezzolla {\it
et. al.} \cite{luci17} have combined the recent gravitational-wave
observation of merging system of binary neutron stars via the
event GW170817 with quasi-universal relations between the maximum
mass of rotating and non-rotating NSs. It is found that the maximum
mass for non-rotating NS should be in the range $2.01\pm0.04\lesssim
M(M_\odot)$$\lesssim2.16\pm0.03$ \cite{luci17}, where the lower limit is
observed from massive pulsars in the binary system \cite{antoni13}. From the
results, we find that the maximum masses for IOPB-I along with FSUGarnet
and G3 EoSs are consistent with the observed lower  bound on the
maximum NS mass. For the IOPB-I parametrization, 
the maximum mass of the NS is $2.15 M_\odot$ and  the radius (without
including crust) of the {\it canonical} mass is 13.242 km which is
relatively larger as compared to the current X-ray observations radii of
range 10.5-12.8 km by N\"attil\"a {\it et. al.} \cite{joonas} and 11-12
km by Steiner {\it et. al.} \cite{steiner10}.  Similarly, FSUGarnet fails
to qualify radius constraint.  However, recently E. Annala {et. al.}
suggest that the radius of a $1.4M_\odot$ star should be in the range
$11.1\leq R_{1.4M_\odot} \leq 13.4$ km \cite{eme17}, which is consistent
with the IOPB-I and FSUGarnet sets.  Furthermore, G3 EoS is relatively
softer at energy density $\cal{E}$$\gtrsim200$ MeV fm$^{-3}$ (see in
Fig. \ref{ep2}), which is able to reproduce  the recent observational
maximum mass of $2.0M_\odot$ as well as the radius of the {\it canonical}
neutron star mass of 12.416 km.

\begin{figure}
        \includegraphics[width=1.1\columnwidth]{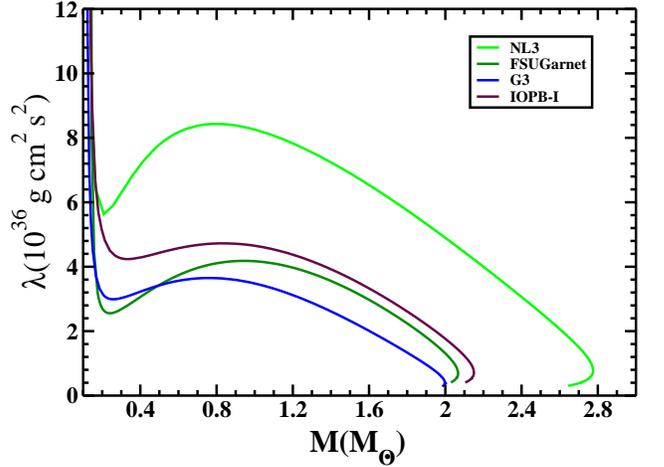}
        \caption{(color online) The tidal deformability $\lambda$ as a function 
of neutron star mass with different EoSs.}
        \label{lamb}
\end{figure}

Now we move to results for the tidal
deformability of the single neutron star as well as binary neutron
stars (BNS) which has been recently discussed in GW170817 \cite{BNS}.
The Eq. (\ref{eq40}) indicates that $\lambda$ strongly depends on the radius
of the NS as well as on the value of $k_2$. Moreover, $k_2$ depends
on the internal structure of the constituent body and directly enters
into the gravitational wave phase of inspiraling BNS which in turn
conveys information about the EoS. As the radii of the NS increases, the
deformation by the external field becomes large as there will be an increase
in gravitational gradient with the simultaneous increase in radius.
In other words, stiff (soft) EoS  yields large (small) deformation
in the BNS system. Fig. \ref{lamb} shows the tidal deformability as
a function of NS mass.  In particular, $\lambda$ takes a wide range
of values $\lambda\sim(1-8)\times10^{36}$ g cm$^{2}$ s$^{2}$ as shown
in Fig. \ref{lamb}. For the G3 parameter set, the tidal deformability
$\lambda$ is very low in the mass region $0.5-2.0M_\odot$ in comparison
with other sets. This is because the star exerts high central pressure
and energy density resulting in the formation of a compact star which is
shown as solid dots in Fig. \ref{ep2}.  However, for the NL3 EoS case,
it turns out that, because of the stiffness of the EoS, the $\lambda$
value is increasing. The tidal deformability of the canonical NS
($1.4M_\odot$) of IOPB-I along with FSUGarnet and G3 EoSs are found to
be 3.191$\times 10^{36}$, 3.552$\times 10^{36}$, and 2.613 $\times 10^{36}$ g cm$^{2}$ s$^{2}$, respectively as shown
in Tables \ref{table4} and \ref{table5}, which are consistent with the
results obtained by Steiner {\it et. al.} \cite{ste15}.

\begin{figure}
        \includegraphics[width=1.1\columnwidth]{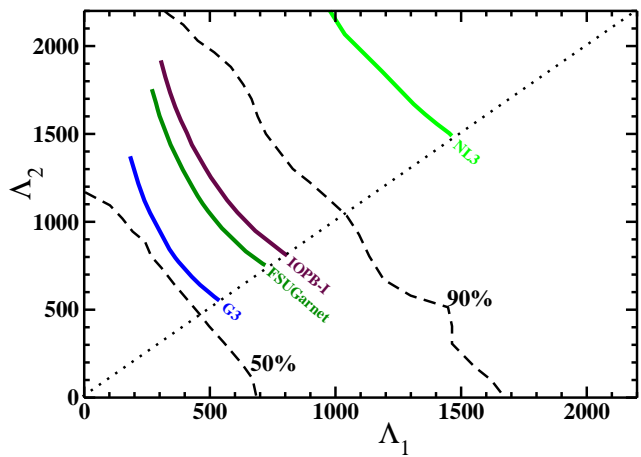}
        \caption{(color online) Different values of $\Lambda$ generated by 
using IOPB-I along with NL3, FSUGarnet and G3 EoSs are compared with the 
$90\%$ and $50\%$ probability contour in case of low-spin $|\chi| \leq 0.05$ 
as given in Fig. 5 of GW170817 \cite{BNS}.}
        \label{tidal}
\end{figure}

Next, we discuss the weighted dimensionless tidal deformability of the 
BNS of mass $m_1$ and $m_2$ and is defined as \cite{BNS,fava14,wade14}:
\begin{eqnarray}
	\tilde{\Lambda}=\frac{8}{13}\Bigg[(1+7\eta-31\eta^{2})(\Lambda_{1}
	\nonumber \\
	+\Lambda_{2})+\sqrt{1-4\eta}(1+9\eta-11\eta^{2})(\Lambda_{1}-
	\Lambda_{2})\Bigg]\; ,
\end{eqnarray}
with tidal correction  
\begin{eqnarray}
	\delta{\tilde\Lambda}=\frac{1}{2}\Bigg[\sqrt{1-4\eta}\Bigg(1-\frac{13272}{1319}\eta+\frac{8944}{1319}\eta^{2}\Bigg)(\Lambda_{1}+\Lambda_{2})\nonumber
	\\
	+\Bigg(1-\frac{15910}{1319}\eta+\frac{32850}{1319}\eta^{2}+\frac{3380}
	{1319}\eta^{3}\Bigg)(\Lambda_{1}-\Lambda_{2})\Bigg].\;
\end{eqnarray}
Here, $\eta=m_{1} m_{2}/M^{2}$  is the symmetric mass ratio, $m_{1}$ and $m_{2}$ are the binary masses,  $M=m_{1}+m_{2}$ is the total mass, $\Lambda_{1}$
and $\Lambda_{2}$ are the dimensionless tidal deformability of BNS, for the
case $m_{1} \geq m_{2}$. Also, we have taken equal and unequal-masses 
($m_1$ and $m_2$) BNS system as it has been done in Refs. \cite{BNS1,david17}. 
The calculated results for the $\Lambda_1$, $\Lambda_2$ and weighted tidal deformability 
$\tilde\Lambda$ of the present EoSs are displayed in Tables \ref{table4} and
\ref{table5}. 
In Fig. \ref{tidal}, we display the different dimensionless tidal
deformabilities corresponding to progenitor masses of the NS. It can be 
seen that the IOPB-I along with FSUGarnet and G3 are in good agreement with 
the $90\%$ and $50\%$ probability contour of GW170817 \cite{BNS}.
Recently, aLIGO/VIRGO detectors have measured the value of $\tilde\Lambda$ 
whose results are more precise than the  results found by considering the
individual value of $\Lambda_{1}$ and $\Lambda_{2}$ of the BNS \cite{BNS}.  
It is noticed that the value of $\tilde\Lambda\leq800$ in the low-spin
case and $\tilde\Lambda\leq700$ in the high spin case within the $90\%$
credible intervals which are consistent with the 680.79, 622.06 and
461.03 of the $1.4M_\odot$ NS binary for the IOPB-I, FSUGarnet and G3
parameter sets, respectively (see Tables \ref{table4} and \ref{table5}
). We also find a reasonably good agreement in the $\tilde\Lambda$
value equal to 582.26 for $1.35M_\odot$ in the G3 EoS, which is
obtained using a Markov Chain Monte Carlo simulation of BNS with
$\tilde\Lambda\approx600$ at signal-to-noise ratio of 30 in a
single aLIGO detector \cite{wade14,fava14}.  Finally, we close this
section with the discussion on chirp mass $\cal{M}$$_c$ and chirp radius
$\cal{R}$$_{c}$ of the BNS system  which are defined as: 
\begin{eqnarray}
	{\cal{M}}_c = (m_1 m_2)^{3/5} (m_1+m_2)^{-1/5}\\ {\cal{R}}_c =
	2 {\cal{M}}_c \;\tilde{\Lambda}^{1/5}.
\end{eqnarray} The precise mass measurements of the NSs have been reported
in Refs. \cite{demo10,antoni13}. However, till now no observation has
been confirmed regarding the radius of the most massive NS. Recently,
aLIGO/VIRGO has measured chirp mass $1.188^{+0.004}_{-0.002}M_\odot$ with
very good precision. With the help of this, we can easily calculate the
chirp radius $\cal{R}$$_{c}$ of the BNS system and we find that the chirp
radius is in the range between $7.867 \leq \cal{R}$$_{c}\leq 10.350 \; $
km for equal and unequal-mass BNS system as shown in Tables \ref{table4}
and \ref{table5}.

\section{Summary and Conclusions}{\label{summary}}

We have built a new relativistic effective interaction for finite nuclei, 
infinite nuclear matter and neutron stars. The optimization has been done
using experimental data for eight spherical nuclei such as binding energy and 
charge radius.  
The prediction of observables such as binding
energies and radii with the new IOPB-I set for finite nuclei are quite good.
The Z=120 isotopic chain shows that the magicity appears at neutron numbers 
N=172, 184, and 198. Furthermore, we find that the IOPB-I set yields
slightly larger values for  
the neutron-skin thickness. This is due to the small strength of the 
$\omega-\rho$ cross-coupling.
For infinite nuclear matter at sub-saturation and supra-saturation densities, 
the results of our calculations agree well with the known experimental data. 
The nuclear matter properties obtained by this new parameter set are:
nuclear incompressibility $K = 222.65$ MeV, symmetry energy
coefficient $J = 33.30$ MeV, symmetry energy slope  $L = 63.6$ MeV,
and the asymmetry term of nuclear incompressibility $K_\tau = 389.46$
MeV at saturation density $\rho_{0} = 0.149$ fm$^{-3}$. In general,
all these values are consistent with current empirical data. The
IOPB-I model satisfies the density dependence of 
symmetry energy which is obtained from the different sets of experimental
data.  It also yields the NS maximum mass to be 2.15$M_\odot$ which is
consistent with the current GW170817 observational  constraint
\cite{luci17}. The radius of the canonical neutron star is 13.24 km
compatible with the theoretical results in Ref. \cite{eme17}.  Similarly,
the predicted values of dimensionless tidal deformabilities are in
accordance with the GW170817 observational probability contour \cite{BNS}.

{\bf Acknowledgments:}
Bharat Kumar would like to take this opportunity to convey special thanks to
P. Landry and Joonas N\"attil\"a, fruitful discussions, and useful suggestions.
Further, Bharat Kumar thankful to Swagatika Bhoi for careful reading of the 
manuscript.

\end{document}